\documentclass[a4paper,11pt]{article}
\pdfoutput=1 
\usepackage{jcappub} 

\usepackage{diagbox}
\usepackage{rotating}
\usepackage{array}
\usepackage{amsmath}
\usepackage{slashed}
\usepackage[normalem]{ulem}
\usepackage{slashed}
\usepackage{booktabs}
\usepackage[pdftex,table]{xcolor}
\usepackage{multirow}
\usepackage{float}
\usepackage{tabularx}
\usepackage{siunitx}
\usepackage{bm}
\usepackage{multirow}
\usepackage{amssymb}

\newcommand{\be}{\begin{equation}}
\newcommand{\ee}{\end{equation}}
\newcommand{\ba}{\begin{array}}
\newcommand{\ea}{\end{array}}
\newcommand{\bea}{\begin{eqnarray}}
\newcommand{\eea}{\end{eqnarray}}

\usepackage{ulem,fancyvrb}
\usepackage{xcolor}

\title{XQC and CSR constraints on 
strongly interacting dark matter 
with spin and velocity dependent 
cross sections}

\author[a]{Yonglin Li,}
\author[a,b]{Zuowei Liu,} 
\author[a]{and Yilun Xue}

\affiliation[a]{Department of Physics, Nanjing University, Nanjing 210093, China}
\affiliation[b]{CAS Center for Excellence in Particle Physics, Beijing 100049, China}

\emailAdd{DZ20220015@smail.nju.edu.cn}
\emailAdd{zuoweiliu@nju.edu.cn}
\emailAdd{mg1722018@smail.nju.edu.cn}

\abstract{Dark matter that interacts strongly with baryons can avoid the stringent dark matter direct detection constraints, because, like baryons, they are likely to be absorbed when traversing the rocks, leading to a suppressed flux in deep underground labs. Such strongly interacting dark matter, however, can be probed by dark matter experiments or other experiments operated on the ground level or in the atmosphere. In this paper we carry out systematic analysis of two of these experiments, XQC and CSR, to compute the experimental constraints on the strongly interacting dark matter in the following three scenarios: (1) spin-independent and spin-dependent interactions; (2) different velocity dependent cross sections; (3) different dark matter mass fractions. Some of the scenarios are first analyzed in the literature. We find that the XQC exclusion region has some non-trivial dependencies on the various parameters and the limits in the spin-dependent case is quite different from the spin-independent case. A peculiar region in the parameter space, where the XQC constraint disappears, is also found in our Monte Carlo simulations. This occurs in the case where the interaction cross section is proportional to the square of the velocity. We further compare our XQC and CSR limits to other experimental constraints, and find that a large parameter space is allowed by various experiments if the dark matter mass fraction is sufficiently small, $f_\chi\lesssim 10^{-4}$.}

\keywords{Strongly interacting dark matter, XQC, Monte Carlo}

\arxivnumber{2209.04387}

\begin{document}

  \maketitle
\flushbottom

\section{Introduction}

Although the experimental evidence of dark matter 
(DM) is overwhelming, 
the property of DM still remains unknown. 
Particle DM is perhaps the most elegant way 
to address the various experimental observations 
on DM. 
Currently, there are several ways to decoding the 
nature of the particle DM: 
DM direct detection, 
DM indirect detection, 
and collider searches. 
In past decades, underground DM direct 
detection experiments have made great 
progress, for example, the Xenon-target DM 
direct detection experiments have probed the 
spin-independent (SI) DM-proton cross section down 
to $10^{-46}$ cm$^2$ \cite{XENON:2018voc,PandaX-II:2020oim}.
However, the interaction cross section 
constrained by such underground experiments have a 
``ceiling'', 
above which DM particles are shielded by the 
rocks on top of the underground labs, 
or even the atmosphere, 
so that they are not constrained by DM 
underground experiments. 
Such DM is often referred to as 
``strongly interacting dark matter'' (SIDM). 
Although SIDM cannot be constrained by underground 
DM direct detection experiments, 
there are ground based DM experiments and 
experiments operated in the atmosphere 
that can probe SIDM:
the XQC rocket experiment \cite{McCammon:2002gb}, 
the CRESST surface run \cite{CRESST:2017ues},
the EDELWEISS-Surf run \cite{EDELWEISS:2019vjv},
the RRS balloon experiment \cite{Rich:1987st},
and satellite experiments \cite{Wandelt:2000ad}.

Although the XQC experiment 
was designed for X-ray spectroscopy, 
it can provide stringent constraints to SIDM 
\cite{Wandelt:2000ad,
Zaharijas:2004jv,Sigurdson:2004zp,Erickcek:2007jv,Mahdawi:2017cxz,Hooper:2018bfw,Mahdawi:2018euy,McKeen:2022poo}. 
The XQC exclusion contours have been computed 
for the spin-independent DM-nucleus cross section 
via Monte Carlo (MC) methods in 
Ref.\ \cite{Erickcek:2007jv}; 
several rather sizeable mass fractions ($f_\chi >0.1$)
have been considered in 
Ref.\ \cite{Erickcek:2007jv}. 
The MC simulations can take into account 
multiple scatterings and thus can provide 
reliable constraints that cannot be easily 
derived via (semi-)analytic methods \cite{Erickcek:2007jv}. 
Recently the lower boundary of the XQC exclusion 
region has been re-analyzed by taking into account 
the DM particles that penetrate the aluminum body 
of the XQC rocket \cite{Mahdawi:2017cxz}, 
leading to a slightly stronger limit 
in the  mass range of $[0.3,100]$ GeV 
than in Ref.\ \cite{Erickcek:2007jv}, 
where DM are assumed to be completely shielded by 
the rocket body.
The lower boundary of the XQC exclusion region 
for different velocity-dependent cross sections 
are computed in Ref.\ \cite{Mahdawi:2018euy}. 
Recently, 
the lower boundary of the XQC exclusion region 
has been re-scaled to estimate the constraints on different DM 
models: 
Ref.\ \cite{Hooper:2018bfw} re-scaled the SI limits 
in Ref.\ \cite{Mahdawi:2017cxz} to obtain  
the limits on spin-dependent (SD) cross sections; 
Ref.\ \cite{McKeen:2022poo} re-scaled the 
constrains in the mass range of $[0.1,100]$ GeV
in Ref.\ \cite{Mahdawi:2018euy} 
to obtain the limits for 
different DM mass fractions; 
Ref.\ \cite{Xu:2020qjk}  re-scaled the results in 
Refs.\ \cite{Mahdawi:2017cxz} 
\cite{Mahdawi:2018euy}
to obtain constraints on Yukawa interactions.

We note that the recent XQC limits 
obtained via analytic methods in 
Ref.\ \cite{Mahdawi:2017cxz} were only for 
a rather narrow mass range of $[0.3,100]$ GeV. 
This is because analytic methods are  
based on the single scattering assumption 
between DM and the detector, which dominates in 
the mass range of $[0.3,100]$ GeV. 
Outside this mass range, as multiple scatterings 
become important, analytic methods are no longer reliable. 
To extend the analysis in Ref.\ \cite{Mahdawi:2017cxz} 
to DM mass of $\sim$0.01 GeV, 
Ref.\ \cite{Mahdawi:2018euy} have used the MC methods, 
which can properly take into account 
multiple scatterings. 
Besides the DM mass $\lesssim 0.3$ GeV, 
we further find that 
the analytical method is not 
a good approximation for DM mass $\gtrsim 10^4$ GeV 
or for mass fraction $\lesssim 10^{-3}$.
Because the analytic method used in 
Refs.\ \cite{Mahdawi:2017cxz,Mahdawi:2018euy}
(as well as the re-scaling method 
used in Refs.\ 
\cite{Hooper:2018bfw,McKeen:2022poo,Xu:2020qjk}) 
cannot be used in the entire XQC exclusion region, 
we use MC method in our analysis. 

In this paper, 
we carry out detailed XQC MC simulations for  
different DM scenarios: 
(1) both SI and SD DM-nucleus cross sections; 
(2) different velocity-dependent DM-nucleus cross sections, 
$\sigma \propto v^n$; 
(3) different DM mass fractions. 
Many XQC MC simulations in this paper have not been analyzed previously, 
such as the upper boundary of the XQC exclusion region 
for the SD and velocity-dependent cross sections. 
We provide the first XQC simulations for the SD 
cross sections, which have been recently estimated 
by rescaling the previous limits in Ref.\ \cite{Hooper:2018bfw}; 
our results for DM with 100\% mass fraction are in good agreement with 
Ref.\ \cite{Hooper:2018bfw}. 
We also provide XQC simulations for very small DM fractions, 
down to $f_\chi = 10^{-5}$. 
Our XQC MC simulations show that  
the upper boundary of the XQC exclusion region depends non-trivially on the mass fraction and on the velocity-dependence. 
We also find that for the velocity-dependent cross section 
with $n=2$, 
there exists a large portion of parameter space 
allowed by the XQC data, which 
cannot be found by rescaling previous results. 

The CSR upper bound on the DM-nucleon cross section 
in the mass range of $[0.1,10]$ GeV in the SI case is given in Ref.\ \cite{CRESST:2017ues}. 
The CSR exclusion region (including both the upper and lower boundaries) 
for DM mass less the 1 GeV in the SI case is analyzed in Ref.\ \cite{Davis:2017noy}. 
The upper boundary of the CSR exclusion region 
in the SI case is extended up to DM 
mass of $10^8$ GeV in Ref.\ \cite{Kavanagh:2017cru}. 
The CSR exclusion regions for both SI and SD cases are computed 
in Ref.\ \cite{Hooper:2018bfw}. 
Recently, MC simulations have been used to more accurately 
determine the CSR exclusion regions 
\cite{Emken:2018run,Mahdawi:2018euy}. 
It is found that the upper boundary of CSR exclusion region for 
DM mass less than 100 GeV in the SI case 
obtained in MC simulations in Ref.\ \cite{Emken:2018run} is in good agreement with 
Refs.\ \cite{Kavanagh:2017cru,Davis:2017noy}, 
but is significantly higher than that in Ref.\ \cite{Hooper:2018bfw}. 
The upper boundary of the CSR exclusion region 
for different velocity dependencies 
and mass fractions for DM mass less than 100 GeV in the SI case
are analyzed with MC simulations in Ref.\ \cite{Mahdawi:2018euy}. 
In our analysis, we compute the CSR exclusion region 
in both the SI and the SD cases  
for different mass fractions. 
In our analysis, we adopt methods that can yield 
consistent results to Ref.\ \cite{Emken:2018run}, 
and thus provide more reliable CSR limits for 
both the SI and the SD cases than Ref.\ \cite{Hooper:2018bfw}.

The rest of the paper is organized as follows. 
In section \ref{sec:DM-model} 
we introduce the three different DM scenarios 
considered in our analysis. 
In section \ref{sec:XQC-infor} 
we introduce the Monte Carlo method to 
compute the XQC constraints. 
We then compute the XQC exclusion regions 
for DM with different velocity-dependent 
cross sections in 
section \ref{sec:XQC-constraint-velocity}, 
for SI and SD DM with different mass fractions in 
section \ref{sec:XQC-constraint-SI} 
and in section 
\ref{sec:XQC-constraint-SD}
respectively. 
In section \ref{sec:CSR} 
we introduce our method to compute the 
constraints on DM from the CSR experiment. 
In section \ref{sec:constraint-mass-fraction} 
we compare the XQC and CSR limits in our analysis 
with various different experimental constraints.
We summarize our findings in section \ref{sec:summary}. 
We further provide some detailed calculations 
in the appendix.

\section{DM models}
\label{sec:DM-model}

In this analysis we consider the following 
three DM scenarios: 
\begin{itemize}
\item 
spin dependence: 
DM interacts with nucleus 
either via the spin-independent cross section 
or via the spin-dependent cross section. 
\item
mass fraction: 
the DM component of interest 
is only a {sub-component}
of the whole DM in the universe. 
\item
velocity dependence: 
the DM-nucleus interaction cross section 
depends strongly on the relative velocity.
\end{itemize}

\subsection{DM-nuclear interaction}
\label{sec:DM-N-interaction}

The interaction between DM and nucleus is usually classified into two different types: 
spin-independent (SI) 
and spin-dependent (SD) interactions. 
For these two interactions, 
the DM-nucleus cross section is related to the 
DM-proton cross section via \cite{Hooper:2018bfw} 
\begin{equation}
\sigma_{\chi N}=\sigma_{\chi p}\left(\frac{\mu_{\chi N}}
{\mu_{\chi p}}\right)^{2} \times\left\{\begin{array}{cc}
{\left[Z+\frac{f_{n}}{f_{p}}(A-Z)\right]^{2}} & \text {(SI)},\\
{\left[\left\langle S_{p}\right\rangle+
\frac{f_{n}}{f_{p}}\left\langle S_{n}\right\rangle\right]^{2} 
\frac{4}{3} \frac{J_{A}+1}{J_{A}}} & (\mathrm{SD}),
\end{array}\right.
\label{eq:DMxsec}
\end{equation}
where 
$\sigma_{\chi N}$ ($\sigma_{\chi p}$) is the DM-nucleus (DM-proton) cross section, 
$\mu_{\chi N}$ ($\mu_{\chi p}$) is the reduced mass of the DM-nucleus (DM-proton) system,
$f_p$ ($f_n$) is the DM-proton (DM-neutron) coupling,
$Z$ ($A$) is the proton (atomic) number of the nucleus $N$, 
$J_A$ is the spin of the nucleus $N$,
and $\left\langle S_{p}\right\rangle$ and $\left\langle S_{n}\right\rangle$ are the average spin of proton and neutron respectively in $N$.
For the SD interaction, only the nucleus with spin\footnote{All nuclei with even mass number have no angular momentum ($I=0$), with the exception of the so-called odd-odd nuclei in which $Z$ and $N$ are both odd \cite{blatt1991theoretical}.}
can participate in the scattering with DM.
See appendix \ref{sec:natural-abundance} 
for the list of the nuclei with spin
considered in this analysis.
We note that the scaling relations given in 
Eq.\ \eqref{eq:DMxsec} may fail for very large 
DM-nucleus 
interaction cross sections 
\cite{Digman:2019wdm} 
\cite{Xu:2020qjk}. 
For the sake of comparison with different 
detector targets, we use the 
scaling relations in  
Eq.\ \eqref{eq:DMxsec} throughout 
our analysis. 
However, one should interpret the very large 
cross section region in a model-dependent manner 
to ensure the validity of the Born approximation 
which is needed to derive the scaling relation 
\cite{Digman:2019wdm} 
\cite{Xu:2020qjk}.

\subsection{Velocity-dependent cross section}

We consider DM models in which 
the DM-proton cross section 
can be parameterized as 
\be
\sigma_{\chi p}=\sigma^0_{\chi p} v^n,
\label{Eq:sigma-vdependent}
\ee
where $\sigma^0_{\chi p}$ is velocity-independent. 
In our analysis, we 
work with phenomenological models and
consider the $n=\{0, 2, -2, -4\}$ cases. 
There are some well-motivated realizations of
the various velocity dependencies.
For example, millicharged DM scatters with 
the nucleus via a t-channel photon, which results in 
an interaction cross section proportional to $v^{-4}$; for recent studies on millicharged DM, see e.g., Refs.\ \cite{Davidson:2000hf,
Dubovsky:2003yn,
Cheung:2007ut,
Feldman:2007wj,
Feldman:2007nf,
McDermott:2010pa,
Cline:2012is,
Dolgov:2013una,
Vogel:2013raa,
Kamada:2016qjo,
Munoz:2018pzp,
Berlin:2018sjs,
Kovetz:2018zan,
Liu:2018jdi,
Liang:2019zkb,
ArgoNeuT:2019ckq,
Plestid:2020kdm,
Ball:2020dnx,
Marocco:2020dqu,
Li:2021kso}. 
The $n=\pm 2$ case can arise in dipole DM models  
\cite{Sigurdson:2004zp}. 
For a variety of 
velocity-dependent cross sections 
in the non-relativistic limit, see e.g.,  Ref.\ \cite{Fan:2010gt}.

\subsection{Mass fraction}

We consider DM models in which SIDM is only a {sub-component} DM. 
Typically, 
the direct detection limits is inversely 
proportional to the DM mass fraction $f_\chi$, 
since the direct detection signal is linearly 
dependent on the number of DM particles. 
However, the upper boundary of the direct detection exclusion 
region has non-trivial dependence on the mass fraction.

\section{XQC constraints on DM}
\label{sec:XQC-infor}

XQC (the X-ray Quantum Calorimetry experiment), 
the rocket experiment originally designed to detect the diffuse X-ray in the range of $60-1000$ eV \cite{McCammon:2002gb}, 
can also place leading constraints to strongly interacting DM \cite{Erickcek:2007jv,Mahdawi:2018euy}. 
XQC was launched in 1999 and collected data 
at 165-225 km above Earth's surface for 100.7 s \cite{McCammon:2002gb,Erickcek:2007jv}. 
The detector of the XQC experiment consists of 36 quantum calorimeters 
with 34 active during the data-taking. 
Each calorimeter consists of a 0.5 mm $\times$ 2 mm $\times$ 0.96 $\mu$m HgTe deposited on a 0.5 mm $\times$ 2 mm $\times$ 14 $\mu$m silicon substrate.
In our Monte Carlo analysis we neglect the effects of HgTe and 
only consider the effects of silicon in the detector.
The FoV of the XQC experiment is 1 sr which is centered at $(\ell,b) = (90^\circ, 60^\circ)$ 
in the galactic coordinate system 
\cite{Erickcek:2007jv}.
There are five filters,  
located at 2 mm, 6 mm, 9 mm, 11 mm and 28 mm above the detectors, 
which are {intended} to block the 
long wavelength radiation for its designed experimental goal \cite{McCammon:2002gb}. 
Each filter consists of a 150 $\rm {\AA}$ aluminum supported on a 1380 $\rm {\AA}$ parylene substrate. 

\subsection{MC simulation on XQC}
\label{sec:MC-XQC}

To compute the XQC constraints on DM,  we carry out MC simulations in the following three steps.

\begin{itemize}

\item 

{\it Step 1:} 
We first select DM particles with velocity in the FoV of XQC, which is 1 sr centered at $(\ell,b) = (90^\circ, 60^\circ)$, 
where $\ell$ ($b$) is the longitude (latitude) in the galactic coordinate system.\footnote{The origin of the galactic coordinate system is the Sun, the x-axis points to the galactic center, 
the y-axis is along the velocity direction of the Sun in the halo frame, 
and the z-axis is along $\hat{z}=\hat{x}\times \hat{y}$.
\label{fn:galactic}
}
We use a truncated Boltzmann distribution \cite{Lewin:1995rx,Essig:2015cda} 
for DM in 
the halo frame
\be
    f(v)=\frac{1}{{N}} e^{-v^2/v_0^2}\theta(v_{\rm esc}- v),
    \label{Eq:v-distribution}
\ee
where $v$ is the DM velocity in the halo frame, $v_0$ is the average velocity of DM, 
$v_{\rm esc}$ is the escape velocity, 
$\theta(v)$ is the Heaviside function, 
and ${N} = \pi v_{0}^{3}  [\sqrt{\pi} \operatorname{Erf}\left(a\right)-2 a e^{-a^{2}} ]$ 
is the normalization factor with $a = v_{\rm esc}/v_0$. 
We take $v_0 = 220$ km/s and $v_{\rm esc} = 587$ km/s \cite{Erickcek:2007jv}.

The total number of DM particles that 
are inside the FoV on the surface of the top filter
during the XQC data-taking is given by 
\be
    N_\chi \simeq 7.32\times 10^8 \, f_\chi \, \frac{\rm GeV}{m_\chi},
\label{Eq:Nchi-filters}
\ee
where we have used 
$S \simeq 7.1$ cm$^2$ 
for the area of the filter. 
See appendix 
\ref{sec:Nchi} for the derivation of Eq.\ \eqref{Eq:Nchi-filters}.

\item

{\it Step 2:} 
We then let DM with the correct velocity traverse the five filters and the detector.  
Each filter consists of a 15 nm Al layer with mass density $\rho_{\rm Al} \simeq 2.7$ g/cm$^3$ and a 138 nm parylene layer with mass density $\rho_{\rm CH} \simeq 1.1$ g/cm$^3$. 
We assume equal composition of $C$ and $H$ in the 
parylene layer. 
The detector is made of Si with mass density $\rho_{\rm Si} \simeq 2.33$ g/cm$^3$. 
The mean free path $\lambda$ of DM in the filter/detector is determined by 
\be
  \lambda = \left( \sum_A n_A \, \sigma_{\chi A} \right)^{-1}, 
\ee
where $n_A$ is the number density of the nucleus $A$, 
and $\sigma_{\chi A}$ is the DM-A cross section. 
The summation is computed on all nuclei in the filter/detector. 
The distance traveled in each step is given 
by $-\ln(1- R)\lambda$, where $R$ is a random number 
between 0 and 1. 
The direction of the motion is given by the DM velocity. 
When there are multiple nuclei to interact with, 
the probability of DM scattering with nucleus A (in each step) is given by 
\begin{equation}
P(A)=\frac{n_{A} \sigma_{\chi A}}{\sum_{B} n_{B} \sigma_{\chi B}}. 
\end{equation}

\item

{\it Step 3:} 
For each given nucleus we compute the DM velocity with two random numbers 
for the polar angle and azimuthal angle in the CM frame. 
See Appendix \ref{sec:scattering:angle} for details. 
In our simulations, for simplicity, we assume 
that DM-nucleus cross sections are 
isotropic in the CM frame. 
Interaction cross sections with non-trivial angular 
distributions are beyond the scope of this analysis.

\end{itemize}

We repeat the simulation steps (1-3) as described above until one of the following conditions is
satisfied: 
(1) The kinetic energy of DM falls below 1 eV, 
or below 29 eV (the lowest energy of XQC data bins) before reaching the detector; 
(2) the energy deposited in the detector exceeds 4000 eV 
(the last XQC data bin has energy $\geq$4000 eV); 
(3) DM does not hit the detector/filter when reaching the detector/filter.

To facilitate the simulation, we use a 8.5 mm $\times$ 4 mm $\times$ 14 $\mu$m 
cuboid to model the detector. 
The radius of the filters is taken to be 1.5 cm, 
as inferred from the Fig.\ 1 of Ref.\ \cite{McCammon:2002gb}. 
For simulations near the lower boundary of the exclusion region, 
we do not consider the effects of the filters,  
because the probability of interacting with filters 
is small for these regions.

Recently, the lower boundary of the 
XQC exclusion region 
has been reanalyzed in the mass region of $[0.3,100]$ GeV 
by Ref.\ \cite{Mahdawi:2017cxz} 
to take into account the probability of DM penetrating the 
aluminum shield (with a thickness of 3.7 cm) of the rocket, 
and in the mass region of $[0.01,100]$ GeV 
by Ref.\ \cite{Mahdawi:2018euy} 
to include the thermalization effects of the silicon detector.  
Because the thickness of the aluminum shield 
is  smaller than the DM mean free path 
on the lower boundary of the XQC exclusion region 
in the mass range of $\sim[0.3,100]$ GeV, 
inclusion of DM penetrating the 
aluminum shield leads to 
a factor of few improvement 
on the XQC limits {in that mass range}
\cite{Mahdawi:2017cxz,Mahdawi:2018euy}. 
However, as one moves outside of the above mass range, 
the thickness of the aluminum shield quickly 
exceeds the DM mean free path 
so that DM can no longer penetrate the shield easily. 
Thus, we have neglected this effect in our 
MC simulations for the entire parameter space. 
As pointed out by Ref.\ \cite{Mahdawi:2018euy}, 
the actual measured energy in the XQC silicon detector 
could be much smaller than the DM nuclear recoil energy. 
However, to our knowledge, 
the thermalization efficiency for silicon 
in the energy range of interest has not been 
measured. 
Thus, in the current analysis, we have not considered 
the effects due to 
the thermalization efficiency for most figures.
To illustrate the effects of the thermalization efficiency, 
we have computed the XQC exclusion region 
adopting the three values of the thermalization efficiency 
in Ref.\ \cite{Mahdawi:2018euy}, 
which are shown in Fig.\ (\ref{fig:XQC-thermal}). 
See Appendix \ref{sec:thermal} for the detailed analysis.

{We next discuss the effects due to 
DM particles that are captured by the Earth. 
As recently pointed out in Ref.\ \cite{Leane:2022hkk},} 
the non-equilibrium component of DM captured by 
the Earth can be significantly larger  
than the equilibrium component 
near the surface of the Earth . 
{We note that both the equilibrium  
and the non-equilibrium components  
are captured and thermalized, 
with the latter possessing a non-zero 
diffusion velocity. 
Because the velocity of the equilibrium component 
is less than the escape velocity of the Earth,  
$v_{\rm esc}^E \simeq 11.2$ km/s {(near the surface)}, 
to produce a recoil energy above 29 eV 
(the energy threshold of XQC), 
the DM mass has to be at least $\sim 40$ GeV. 
At such large DM mass, the density of 
the equilibrium component near 
the Earth surface is much smaller than 
the local DM density \cite{Leane:2022hkk}.} 
To estimate the effects of the 
non-equilibrium component of DM 
on our XQC analysis, 
we use Eq.\ \eqref{Eq:energy-loss} to calculate the 
DM velocity at the XQC detector. 
Following Ref.\ \cite{Leane:2022hkk}, 
we consider the atmosphere up to 180 km.  
To simplify the calculation, we assume 
that the atmosphere is only  
composed of oxygen, 
and has a temperature of $T= 700$ K 
and a density of $\rho=9.75\times 10^{-10}$ kg/m$^3$ \cite{DZIEWONSKI1981297}, 
at the altitude of 165-180 km. 
The condition that DM is captured and thermalized by the Earth is $v_\chi^f\leq v_{\rm esc}^E= 11.2$ km/s and $v_\chi^f\leq v_{\rm th}= \sqrt{8T/\pi m_\chi}$ where $v_{\rm th}$ is the thermal velocity. 
We find that  
at the altitude of 165 km, 
the cross section needed 
to capture  
and thermalize DM 
are much larger than the upper boundary of XQC exclusion region. 
For example, 
the minimal cross sections to capture DM 
and to thermalize DM for a 40 GeV DM, 
are found to be 
$\sigma_{\chi p}\sim 7.5\times10^{-21}$ cm$^2$ and 
$\sigma_{\chi p}\sim 1.5\times10^{-20}$ cm$^2$
respectively.
{Thus, inside the XQC exclusion region, 
DM is not able to be captured and thermalized 
at the upper atmosphere due to its 
low density.}

\subsection{XQC data and $\chi^2$ analysis}
\label{sec:chi2-analysis}

We use the XQC data from table 1 of Ref. \cite{Erickcek:2007jv}. 
Following Ref.\ \cite{Erickcek:2007jv} 
we carry out the following $\chi^2$ analysis to compute the XQC exclusion region
\be
\chi^{2} = \underset{i}
 \sum \left\{\frac{\left(E_{i}-U_{i}\right)^{2}}{E_{i}} 
 \text { with } U_{i}<E_{i}\right\}, 
\label{Eq:chi2}
\ee
where 
$i$ denotes the $i$-th data bin, 
$E_i$ is the number of simulated events
and $U_i$ is the number of events in the XQC data. 
To compute $E_i$ we also take into account the 
XQC detection efficiency: The detection efficiency of XQC is $f = 0.3815 \, (0.5083)$ for the first (second) energy bin, and $f=1$ for the other eleven high energy bins \cite{Erickcek:2007jv}.
Here we use the one-side $\chi^2$ because the background of the XQC experiment is unknown; thus we only consider the cases where the events that DM generate exceed the XQC data. 
The 90\% CL exclusion region is then obtained by 
requiring $\chi^2=2.71$.

\section{XQC constraints on velocity-dependent cross section}
\label{sec:XQC-constraint-velocity}

\begin{figure}[tbp]
    \centering
    \includegraphics[width=0.9\textwidth]{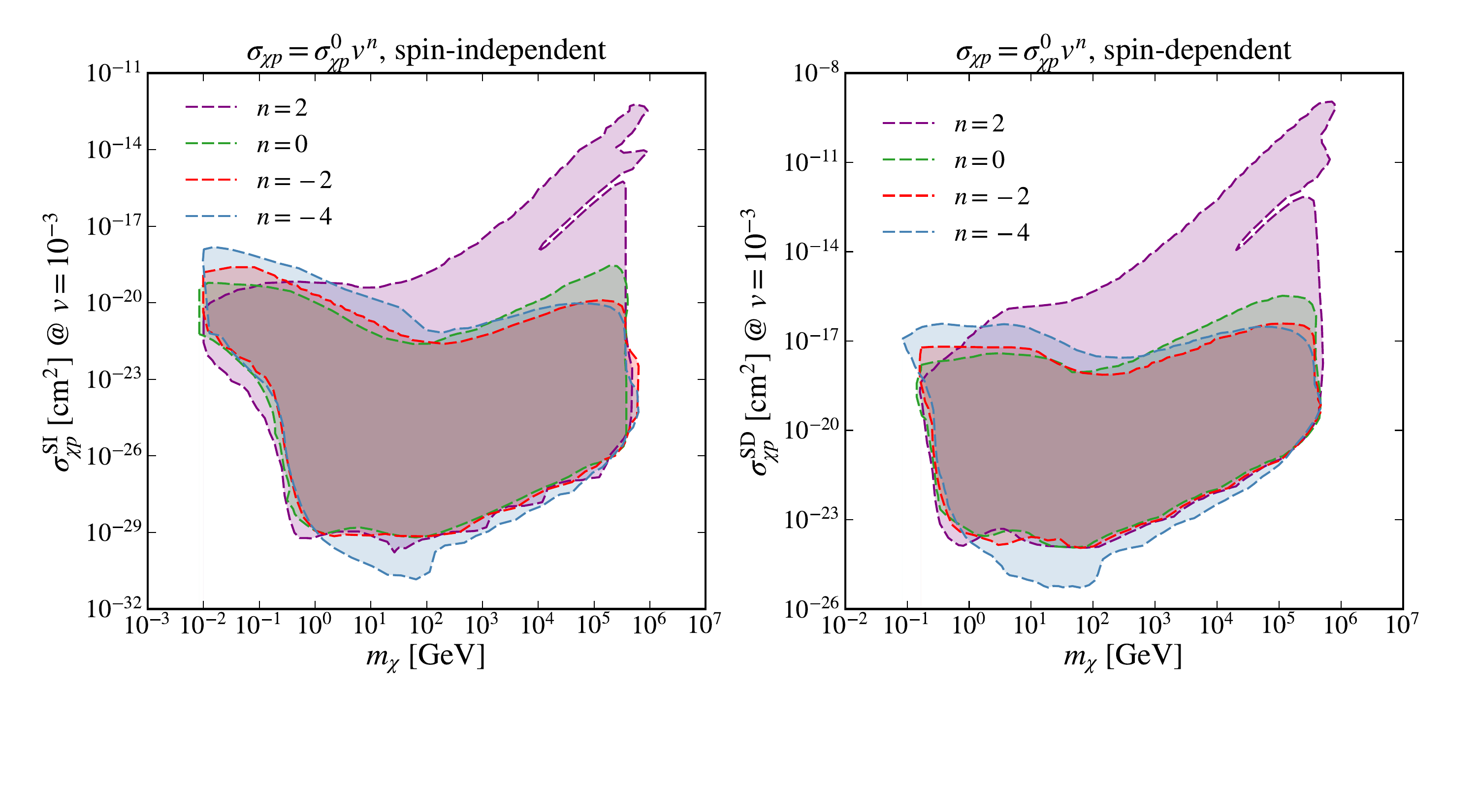}
    \caption{The XQC exclusion region 
    at the 90\% CL in the parameter space 
    spanned by the DM-proton cross section 
    $\sigma_{\chi p}$ evaluated at $v=10^{-3}$ and the DM 
    mass $m_\chi$, 
    both in the SI case ({\it left}) 
    and in the SD case ({\it right}). 
    Different velocity-dependent DM-p cross sections in the form of $\sigma_{\chi p} = \sigma_{\chi p}^0 v^n$,
    where $n=\{0,2,-2,-4\}$, 
    are indicated by different color-shaded regions.}
    \label{fig:XQC-v}
\end{figure}

We compute XQC constraints on 
DM models in which the DM-proton cross section 
are velocity-dependent and 
can be parameterized as in Eq.\ \eqref{Eq:sigma-vdependent}.
Fig.\ (\ref{fig:XQC-v}) shows 
the XQC constraints on DM models 
with $n=\{0, 2, -2, -4\}$ 
both for SI interaction 
and for SD interaction. 
To compare XQC constraints for different $n$ values, 
we computed the excluded DM-proton cross section at $v=10^{-3}c$.
Fig.\ (\ref{fig:XQC-v}) shows that most of the exclusion regions 
for the four different velocity dependencies overlap with 
each other. 
However, there are some noticeable differences 
among them: 
(i) the upper boundary of the exclusion region for heavy DM mass increases with $n$,  with the $n=2$ case significantly higher than other cases;
(ii) there exists a band-shape parameter space allowed by XQC 
in the $n=2$ case that is inside the XQC region, 
which 
can be parameterized as %
\be
    \log_{10}\left[{\sigma_{\chi p} (v=10^{-3})\over{\rm cm}^2}\right] =  k\log_{10}\left[{m_{\chi}\over{\rm GeV}}\right]-b, 
    \label{eq:band}
\ee
where $k \simeq 2 $ and $b \simeq -26 \, (-22.6)$ 
for the SI (SD) case.
(iii) in the $n=-4$ case, 
the lower boundary of the exclusion region 
is lower than other cases, 
and the upper boundary for low mass is 
higher than other cases.

We note that on the lower boundary of the XQC exclusion region, 
the scattering probability of DM with the detector is typically 
very small so that DM can only scatter with the detector once, 
and the scattering between DM and the 5 filters can be 
ignored (except for the very small mass region). 
However, 
on the upper boundary of the XQC exclusion region, 
the interaction cross section is so large 
that the scattering between DM and the 5 filters 
becomes very important.

We note that, for the $f_\chi=1$ case,
the upper boundary and a significant portion 
of the XQC exclusion 
region have been ruled out 
by the CMB and Lyman-$\alpha$ constraints; 
see Fig.\ (\ref{fig:exclusion}) for the comparison. 
Nonetheless, it is of importance to study the 
XQC constraints on various velocity-dependent cross sections, 
because CMB and Lyman-$\alpha$ constraints 
are derived based on physics processes in the early universe, 
but XQC constraints are obtained in the current time. 
The underlying assumption of the comparisons between 
different experimental limits in Fig.\ (\ref{fig:exclusion}) 
is that the particle identity and the mass fraction 
of DM do not change 
from the CMB epoch to the current time, 
which may not be true in a real cosmology 
model.

\subsection{The $n=2$ case}

For the $n=2$ case, there exists certain parameter space 
in which the kinetic energy of the DM particle after 
traversing the five filters is localized in a narrow 
energy range; 
see appendix \ref{sec:energy-loss}
for more detailed discussions 
on the final kinetic energy distribution of DM. 
If the narrow energy range of the DM final kinetic energy 
falls inside $[2505,4000]$ eV, the disregarded XQC energy 
range, then there is no XQC constraints, which 
results in the band-shape parameter space 
in the $n=2$ case. 
If the narrow energy range of the final kinetic energy 
falls inside any other data bin, it can result in a much 
larger $\chi^2$ than the case where the events are 
more evenly distributed among many data bins. 
This then gives rise to stronger constraints 
near the upper boundary of the 
XQC exclusion region in the $n=2$ case. 
We note that the two features of the XQC exclusion region 
in the $n=2$ case should also appear in $n>2$ cases.

\subsection{The lower boundary in the $n=-4$ case}

We next discuss the lower boundary in the $n=-4$ case. 
Because the DM-p cross section has a significant 
dependence on the velocity such that the cross section increases 
significantly when the velocity decreases. 
On the lower boundary of the exclusion region, 
the scattering probability of DM with 
the XQC detector is small so that
DM particles with low velocity have more contributions to 
the interactions between DM and the detector. 
When evaluating the cross section at relatively large 
velocity, such as $v=10^{-3}$ as done in Fig.\ (\ref{fig:XQC-v}), 
one should obtain a smaller cross section 
than the low velocity. 
This explains why the lower boundary of the XQC exclusion 
region in the $n=-4$ case is lower than other cases.

\subsection{The upper boundary in the $n=-4$ case}

We next discuss the upper boundary in the $n=-4$ case. 
On the upper boundary of the XQC exclusion 
region, the interactions between DM and the 5 filters 
play a dominant role in determining the constraints. 
In our analysis, we assume an isotropic scattering 
cross section in the CM frame. 
For light DM ($m_\chi \ll m_N$), 
because the lab frame is nearly the same as the CM frame, 
the scattering cross section between DM and the filters 
in the lab frame is also isotropic. 
Thus, large-angle scatterings can occur quite often
and light DM are easily scattered outside of the 
XQC FoV during interactions with the filters. 
On the contrary, for heavy DM ($m_\chi \gg m_N$),
the lab frame is no longer the CM frame, 
and the scattering cross section 
is mostly forward in the lab frame. 
Thus, heavy DM particles mainly 
lose their kinetic energies via interactions 
with the filters, 
but keep the direction of motion nearly fixed.

We next compare the XQC upper boundary 
for the $n=0$ and $n=-4$ cases. 
As shown in both panel figures 
in Fig.\ (\ref{fig:XQC-v}), 
for DM mass 
below (above) $\sim 10^3$ GeV, 
the upper boundary in the $n=-4$ case is 
higher (lower) than the $n=0$ case. 
Below we discuss two benchmark DM 
masses: 1 GeV and $10^5$ GeV, 
to provide some qualitative 
understandings of this phenomenon.
We first note that the number of DM particles of 1 GeV  
is $10^5$ times larger than DM of $10^5$ GeV, 
since $n\propto m_{\chi}^{-1}$. 
For DM of 1 GeV, 
most DM particles are scattered outside of the XQC FoV 
via large-angle scatterings in the lab frame. 
On the contrary, for DM of $10^5$ GeV, 
most DM particles penetrate the filters to reach 
the detector (if they initially aim right at the detector). 
The upper boundary of the heavy DM is determined 
by the filters such that  
the DM kinetic energy after multiple scatterings with 
the filters is below the XQC threshold. 
Due to the different interactions between the filters 
and the DM, the mean free path of 
DM of 1 GeV is many times larger 
than DM of $10^5$ GeV.

We discuss the upper boundary for the low DM mass. 
In the $n=0$ case, the stopping effects are 
more or less 
the same for all DM velocity, since the interaction 
cross section is velocity independent. 
However, for the $n=-4$ case, the stopping effects 
are significantly weakened for the high energy 
DM particles so that DM particles 
with high velocity has a 
higher probability to penetrate the 5 filters 
than those with low velocity. 
Thus, for the exclusion cross section 
(evaluated at $v=10^{-3}$) at the 
upper boundary of the $n=0$ case, 
DM particles of high velocity are absorbed by 
the filters in the $n=0$ case, 
but these DM particles in the 
$n=-4$ case can penetrate the filters. 
Thus, one needs a larger cross section to stop 
these DM particles of high velocity in the $n=-4$ case. 
Note DM particles that penetrate the filters 
on the upper boundary are likely to be absorbed 
by the detector, since 
the detector is about 100 times thicker than 
the 5 filters.

The situation for the heavy DM is a little complicated. 
Most DM particles penetrate the filters 
with reduced velocities; 
for the $n=-4$ case, the DM of high velocity 
receives a relatively smaller reduction on the 
velocity than the $n=0$ case. 
Because the DM mass is very large, 
the kinetic energies of most DM particles 
are large such that they would result in 
events in the last XQC data bin, $E>4000$ eV, 
since most of the kinetic energy is absorbed by 
the detector. 
For the $n=-4$ case, because the interaction cross section 
at low DM velocity is much larger, 
the number of DM particles with low velocity 
are more easily to lose energy leading to 
more (less) events below (above) the XQC threshold. 
Therefore, with the same cross section, 
there are more events in the last XQC data bin 
in the $n=0$ case than the $n=-4$ case, 
for DM of $10^5$ GeV. 
One then needs a larger cross section 
to reduce the velocity of the DM in the 
$n=0$ case in the heavy DM region.

\section{XQC constraints on SI cross sections}
\label{sec:XQC-constraint-SI}

\begin{figure}[tbp]
    \centering
    \includegraphics[width=0.45\textwidth]{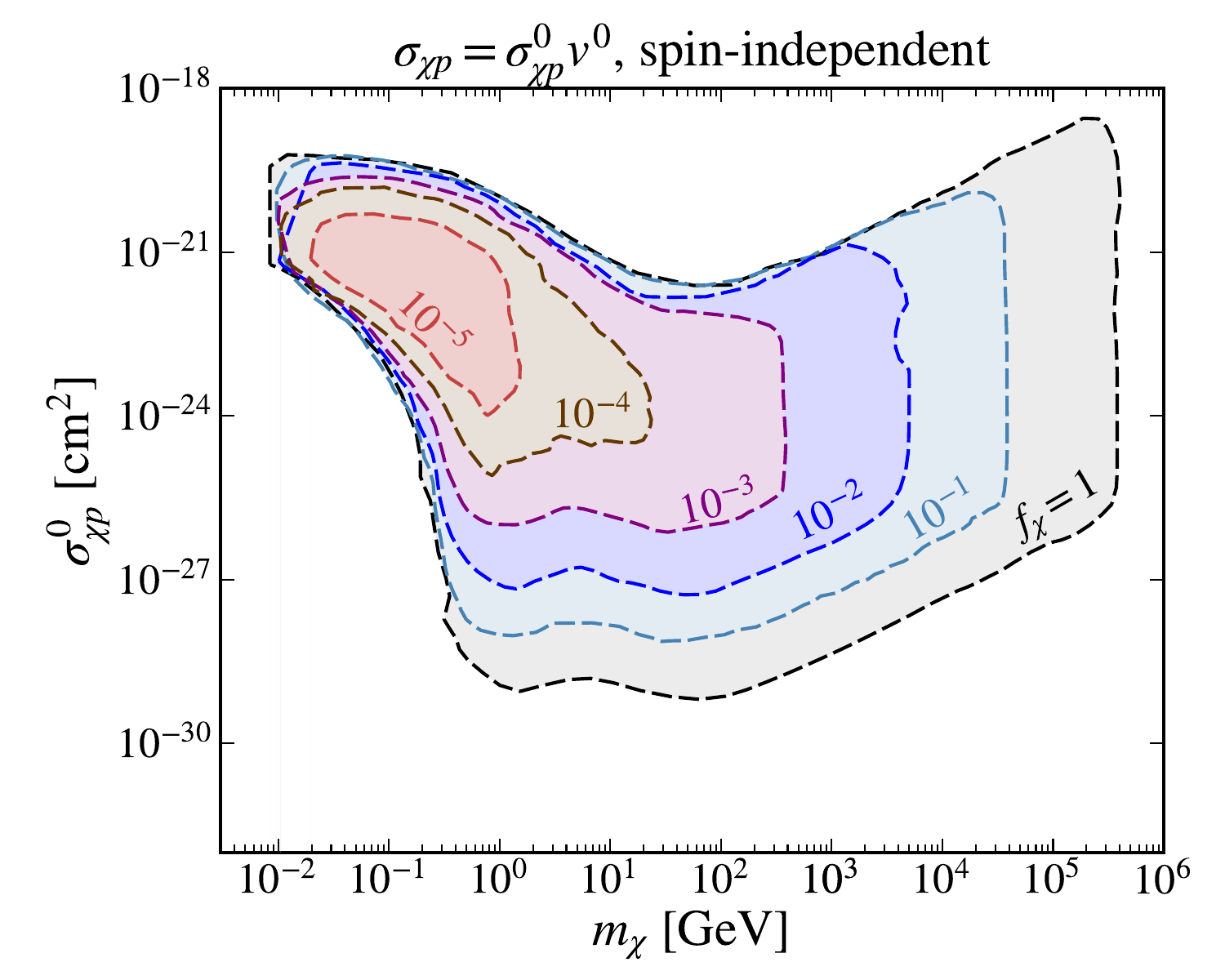}
    \includegraphics[width=0.45\textwidth]{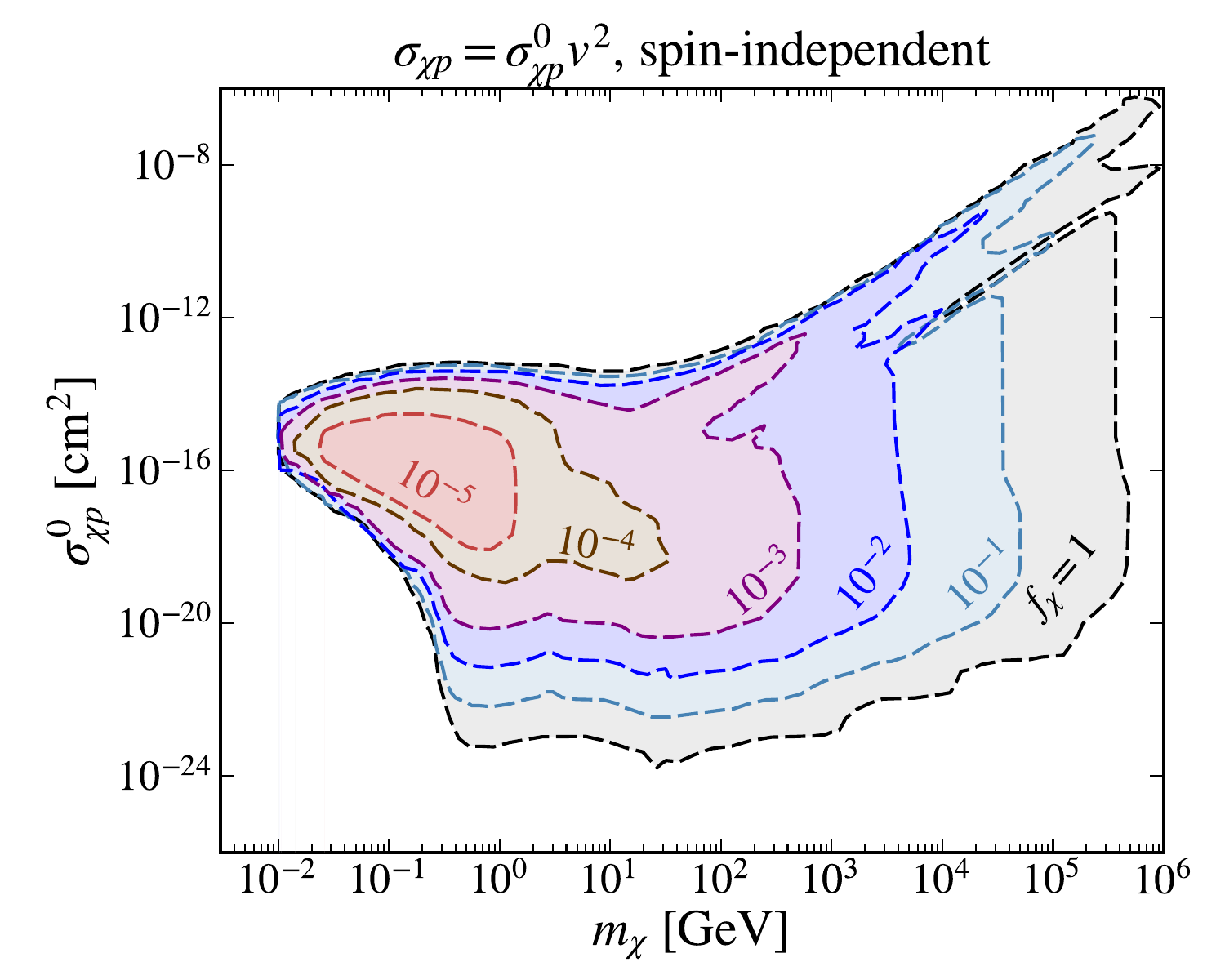}
    \includegraphics[width=0.45\textwidth]{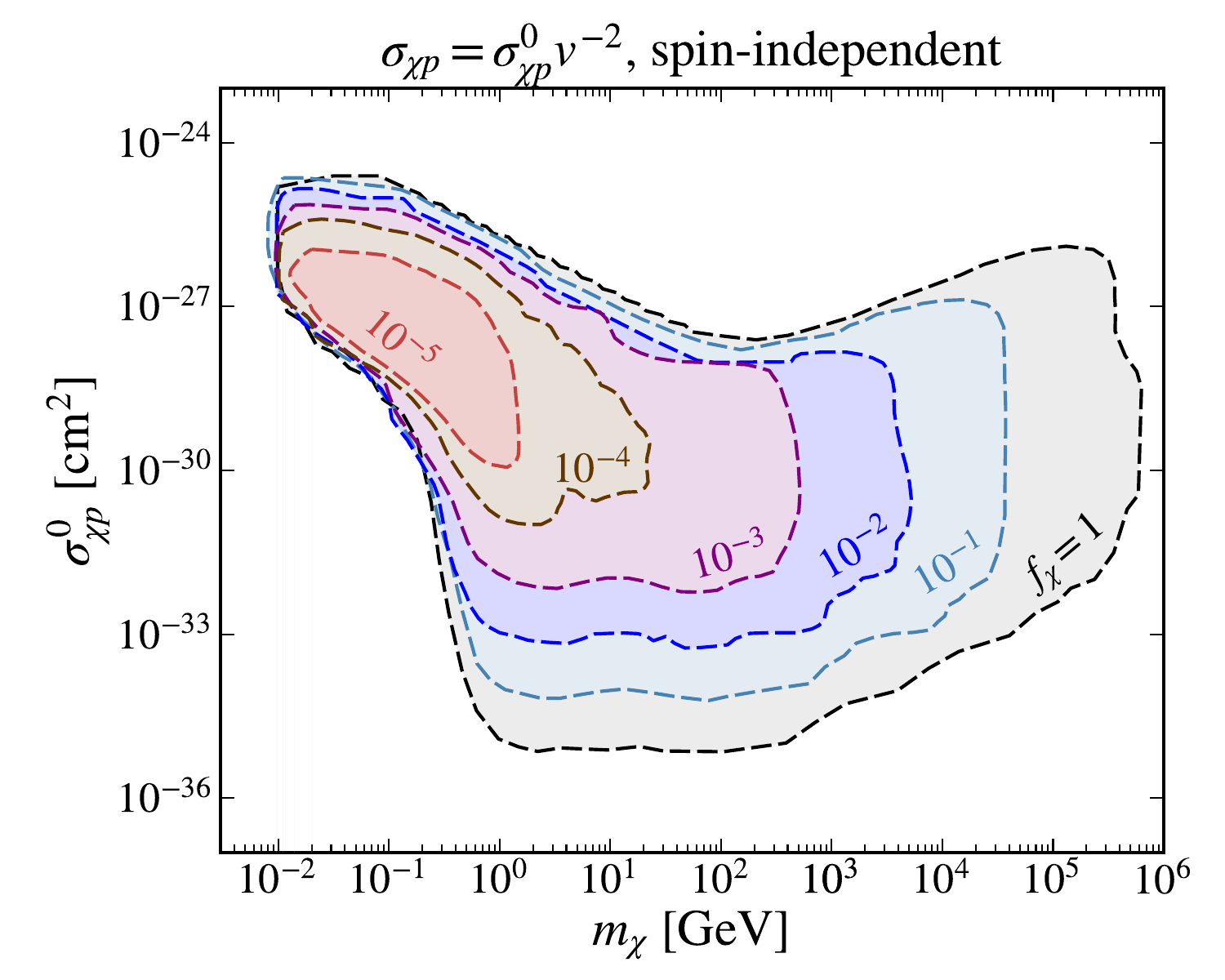}
    \includegraphics[width=0.45\textwidth]{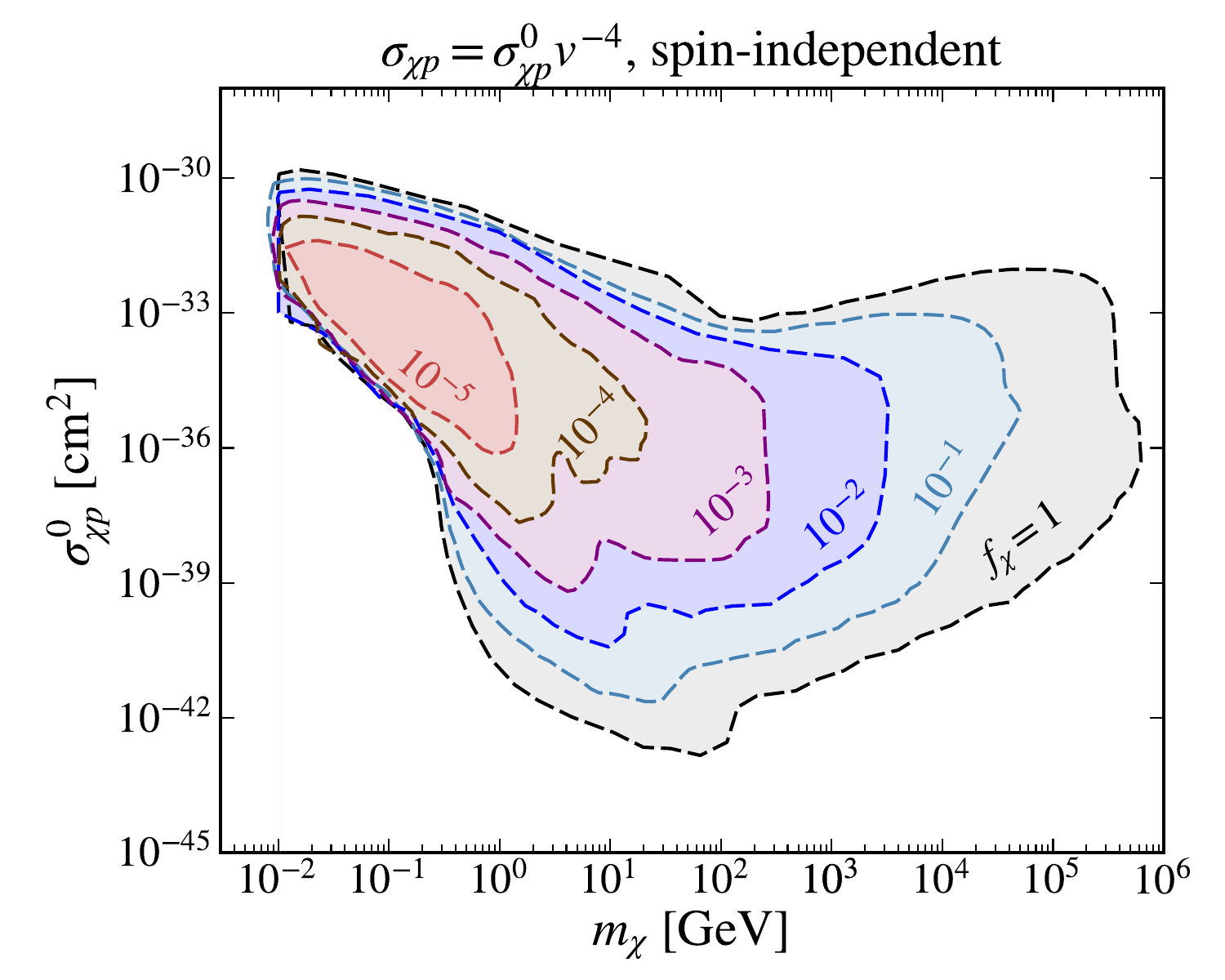}
     \caption{The XQC exclusion region at the 90\% CL 
     in the SI case for 
     different velocity-dependent DM-p cross sections in the form of $\sigma_{\chi p}=\sigma_{\chi p}^0v^n$: 
$n=0$ (upper-left),
$n=2$ (upper-right),
$n=-2$ (lower-left),
$n=-4$ (lower-right).
In each figure, different mass fractions are 
considered: 
$f_\chi=1$ (gray), 
$f_\chi=10^{-1}$ (steel-blue), 
$f_\chi=10^{-2}$ (blue), 
$f_\chi=10^{-3}$ (purple), 
$f_\chi=10^{-4}$ (brown), 
$f_\chi=10^{-5}$ (red).}
    \label{fig:XQC-mf-SI}
\end{figure}

We further compute the XQC exclusion region 
on the DM-proton cross section with different mass fractions. 
Fig.\ (\ref{fig:XQC-mf-SI}) shows the XQC exclusion 
regions for the following mass fractions: 
$f_\chi=\{1,\,10^{-1},\,10^{-2},\,10^{-3},\,10^{-4},\,10^{-5}\}$. 
The XQC exclusion region shrinks as the mass fraction 
decreases. 
The lower boundary and the right boundary of the XQC exclusion 
region have a strong dependence on the mass fraction $f_\chi$, 
whereas the upper boundary and the left boundary have 
a weak dependence on the mass fraction $f_\chi$.

As shown in Fig.\ (\ref{fig:XQC-mf-SI}), 
in the case where $f_\chi = 10^{-5}$, 
all the XQC exclusion regions become very small, 
which are expected to disappear if the mass fraction 
goes below $f \simeq 10^{-6}$. 
This is due to the fact that the number of 
DM particles inside the FoV is significantly small 
for such a small mass fraction so that there are not 
enough signal events to be excluded by XQC even if 
all DM particles are absorbed by the detector. 
The expected 
number of the DM particles that are inside the FoV of the XQC 
detector is 
(see appendix \ref{sec:Nchi} for details)
\be
    N_\chi \simeq 3.5 \times 10^7 f_\chi \, \frac{\rm GeV}{m_\chi},
    \label{Eq:Nchi-detector}
\ee
where we have used the area of 0.34 cm$^2$ for the XQC detector. 
Thus if $f_\chi \lesssim 10^{-6}$, 
$N_\chi \lesssim 35\,({\rm GeV}/m_\chi)$; 
since the total number of events in the XQC data is 
$\mathcal{O}(500)$, there is no constraint 
when $f_\chi \lesssim 10^{-6}$.

For the lower boundary of the XQC exclusion region, 
the DM-p cross section $\sigma_{\chi p}^0$ 
on the lower boundary
is inversely proportional to the mass fraction $f_\chi$. 
This is because on the lower boundary, 
for a significant mass fraction such that the 
expected DM events $N_\chi$ is much larger than ${\cal O}(500)$,  
the interaction probability between DM and the detector 
is very small,
and it is a good approximation to assume that 
DM only interacts with the detector once when traversing the 
detector.
Under the single-scattering assumption, 
the event rate on the lower boundary is thus proportional to 
the product of the interaction cross section and the mass fraction, 
$R \propto \sigma_{\chi p}f_\chi$.
Thus the lower boundary of the XQC exclusion region is
inversely proportional to the mass fraction $f_\chi$.

The right boundary of the XQC exclusion region 
is nearly vertical (namely independent of 
the cross section) for many cases, 
as shown in Fig.\ (\ref{fig:XQC-mf-SI}). 
This truncation arises because
on the right-hand side of the vertical boundary, 
the number of DM particles 
that enter the XQC detector becomes 
less than the number of events observed 
in the last bin of the XQC data.
The location of the right boundary, 
namely the DM mass, moves with the mass fraction 
such that the ratio $f_\chi/m_\chi$ stays unchanged. 
This is due to the fact that the number of DM 
particles in the halo is inversely proportional 
to the DM mass. 
Thus the total number of DM particles remains constant
on the right boundary if 
{$f_\chi/m_\chi$} is fixed. 
In the $f_\chi =1$ case, 
the DM mass is in the range of $10^{5-6}$ GeV 
on the right boundary. 
For such a large DM mass, the energy deposited by 
DM particles in the detector is typically in the last
bin of the XQC data, namely the total deposit energy $E_R \geq 4$ keV.
Thus the right boundary of the XQC exclusion region
is primarily determined by the number of events 
in the last bin of the XQC data, which is $N=60$;
this is valid for $f_\chi \gtrsim 10^{-3}$, 
when $m_\chi\gtrsim 100$ GeV.

In the literature it is often assumed that 
the upper and/or the left boundary of the XQC exclusion 
region do not change when the mass fraction decreases. 
However, as shown in Fig.\ (\ref{fig:XQC-mf-SI}), 
the upper boundary and the left boundary do change 
as the mass fraction decreases. 
We note that the changes on the 
upper boundary and on the left boundary 
are less significant than 
the lower/right boundary. 
This is because on the left boundary the limit is 
set by requiring the kinetic energy 
of each DM particle to be below the XQC threshold, 29 eV;
on the upper boundary the limit is 
set by requiring each DM particle to be absorbed
by the filters.

\section{XQC constraints on SD cross sections} 
\label{sec:XQC-constraint-SD}

In this section, we calculate the XQC exclusion region 
in the SD case. 
In the XQC filters, the nuclei with spin include  
$^{27}$Al, $^{13}$C, and $^1$H; 
in the detector, the nucleus with spin is $^{29}$Si. 
The natural abundance of 
$^{27}$Al and $^1$H are about 100\%, 
and the natural abundance of $^{13}$C and $^{29}$Si 
are 1.1\% and 4.7\% respectively \cite{lodders1998planetary}.

\begin{figure}[tbp]
    \centering
    \includegraphics[width=0.45\textwidth]{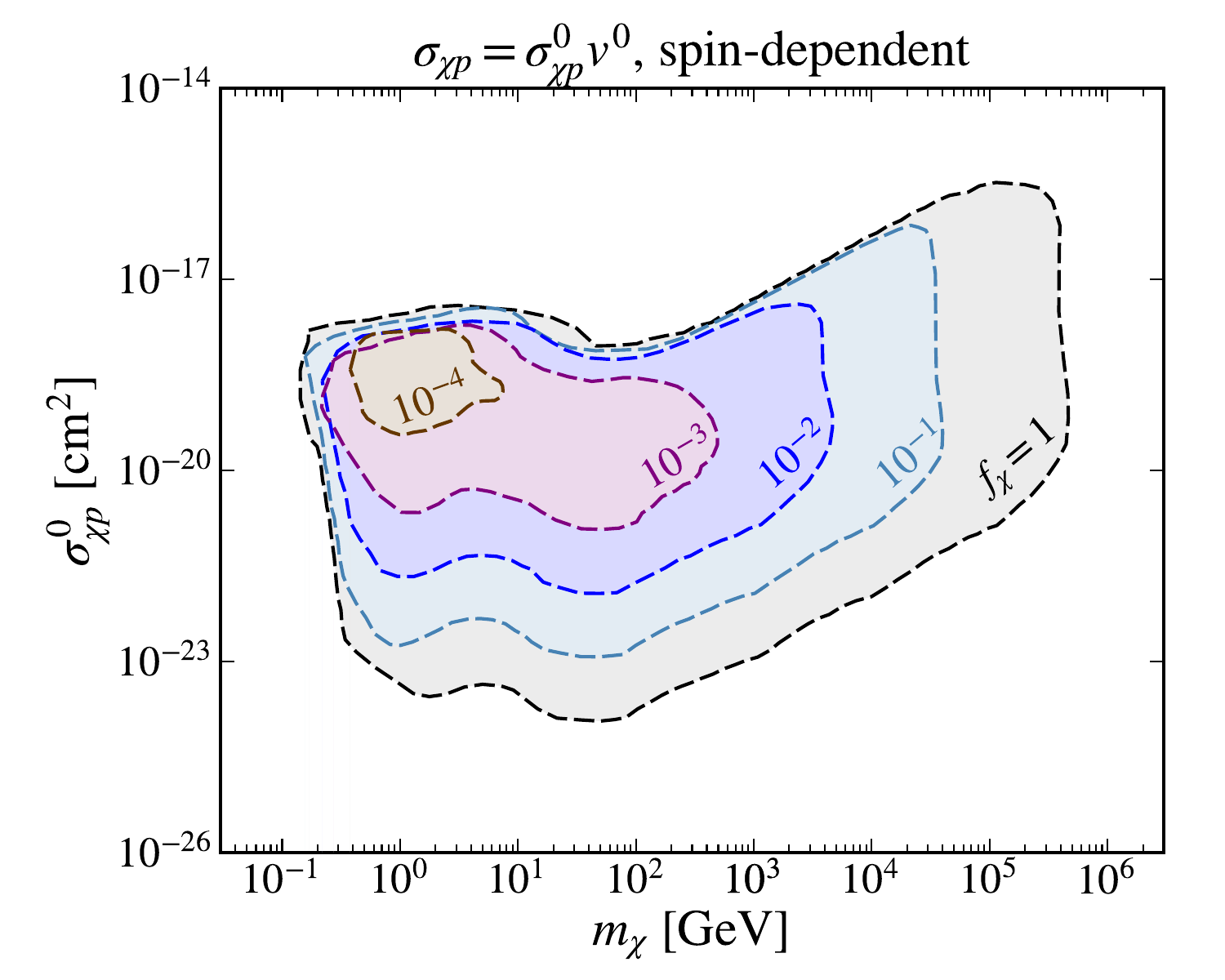}
    \includegraphics[width=0.45\textwidth]{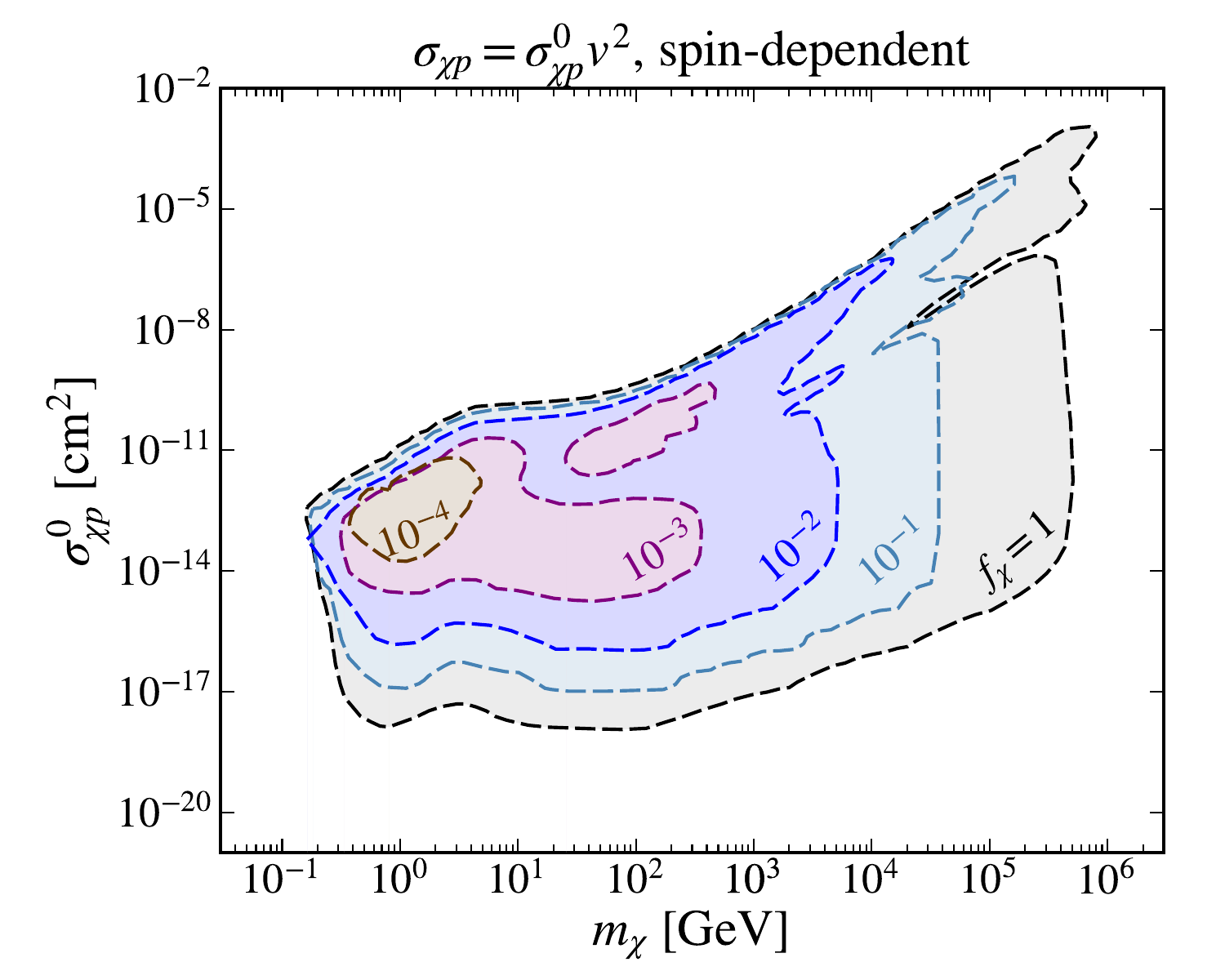}
      \includegraphics[width=0.45\textwidth]{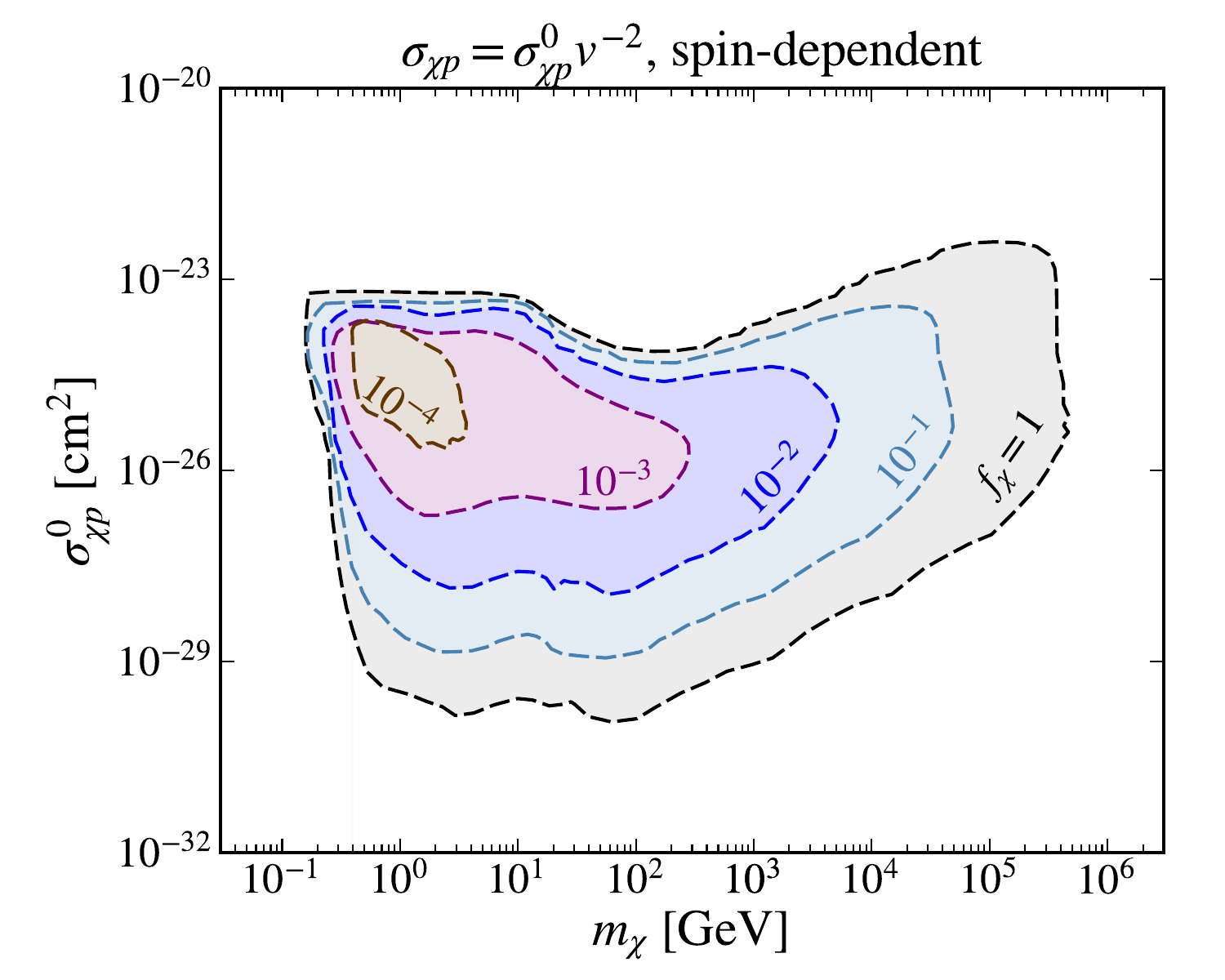}
    \includegraphics[width=0.45\textwidth]{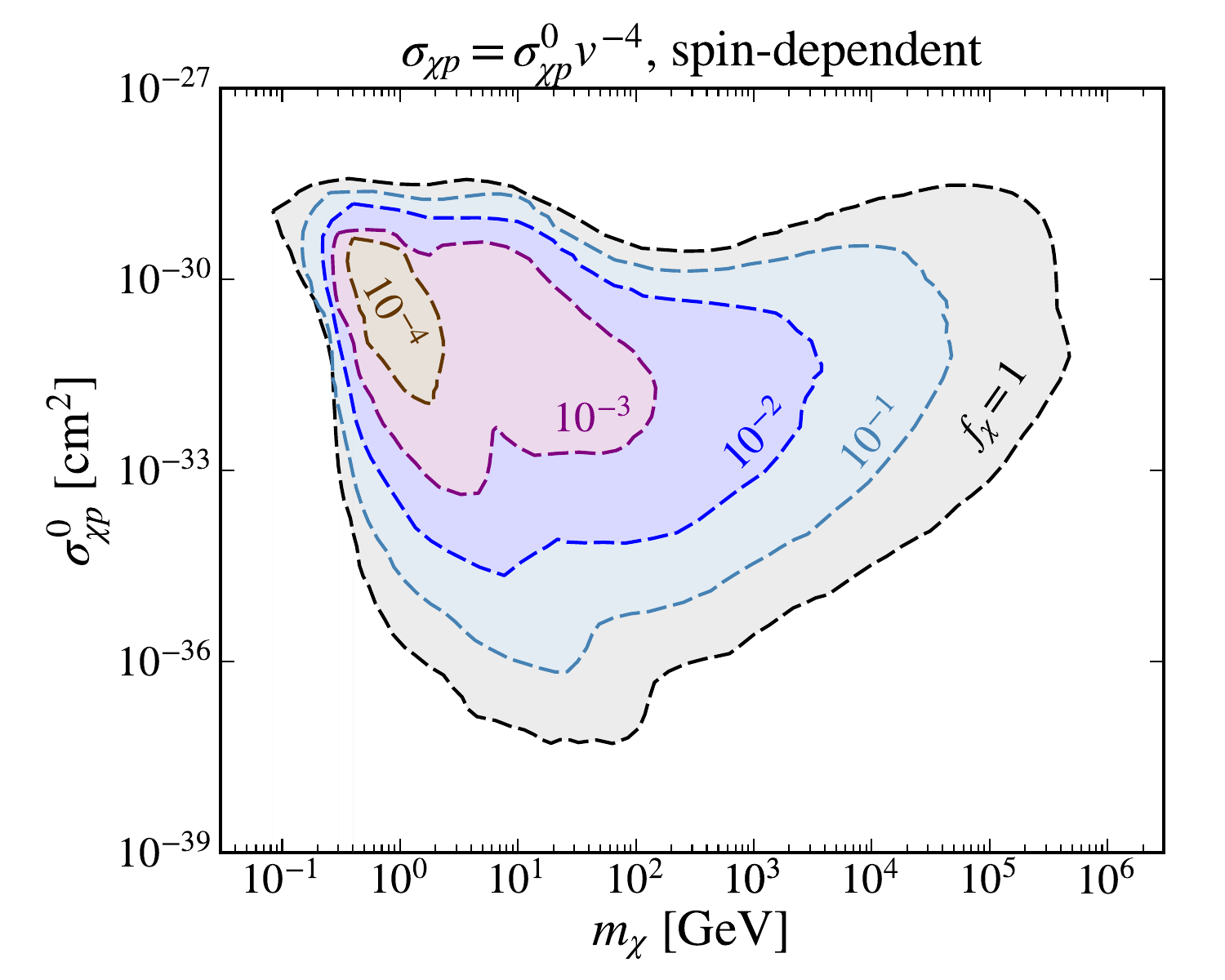}
    \caption{
    The XQC exclusion region at the 90\% CL 
     in the SD case for 
different velocity-dependent DM-p cross sections in the form of $\sigma_{\chi p}=\sigma_{\chi p}^0v^n$:
$n=0$ (upper-left),
$n=2$ (upper-right),
$n=-2$ (lower-left),
$n=-4$ (lower-right).
In each figure, different mass fractions are 
considered: 
$f_\chi=1$ (gray), 
$f_\chi=10^{-1}$ (steel-blue), 
$f_\chi=10^{-2}$ (blue), 
$f_\chi=10^{-3}$ (purple), 
$f_\chi=10^{-4}$ (brown).
}
    \label{fig:XQC-mf-SD}
\end{figure}

Fig.\ (\ref{fig:XQC-mf-SD}) shows the XQC exclusion regions 
in the SD case 
with different mass fractions 
and different velocity-dependent cross sections. 
The lower boundary of the SD case is about $10^5$ times larger than the SI case, as shown in Fig.\ (\ref{fig:XQC-mf-SI}); 
the shape of the lower boundary in the SD case is similar to the SI case. 
This is because on the lower boundary 
the probability of DM scattering with the detector is small 
and the probability of DM scattering with the filters is negligible. 
This rescaling of the limits are then deduced by 
the ratio between the SI and SD cross sections given in 
Eq.\ \eqref{eq:DMxsec} 
and the natural abundance of $^{29}$Si, 
which together give rise to a factor of $\sim 10^5$. 
We note that the shape of the XQC exclusion region on the upper boundary 
in the SD case is also quite different from the SI case: 
for example in the $n=0$ case, 
the upper boundary at small DM mass is much smaller than heavy DM mass 
in the SD case, 
whereas in the SI case, they are comparable. 
We also note that the XQC exclusion region disappears 
when $f_\chi\lesssim10^{-5}$.

\section{Constraints from CSR  experiment}
\label{sec:CSR}
\label{sec:CSR-constraint}

CSR (CRESST surface run) 
is a ground-based experiment 
which
used the cryogenic detector to detect the DM signal 
\cite{CRESST:2017ues}. 
The CSR detector is made of Al$_2$O$_3$, which has a size of $5\times5\times5$ mm$^3$ and a mass of 0.49 g. 
The CSR data consists of 511 events 
in a net live-time of 2.27 h 
which corresponds to a net exposure of 0.046 g-day. 
We compute the CSR exclusion region on DM models 
with different mass fractions for both SI and SD interactions \cite{CRESST:2017ues}. 
Below we describe our calculation on 
the upper boundary and 
the lower boundary of the CSR exclusion region.

\subsection{The lower boundary of the CSR exclusion region}

For the lower boundary we first calculate the event rate via 
\be
\frac{{d} R}{{d} E_R}=
\frac{\rho_\chi }{m_{{\chi}}}
\sum _{A} N_{A} \int {d}^{3} v \, v\, f(v)
\frac{{d} \sigma_{\chi A}}{{d} E_R},
\label{Eq:SDeventrate}
\ee
where $R$ is the event rate, 
$E_R$ is the recoil energy, 
$\rho_\chi$ is the local DM mass density,
$A$ denotes different nuclei in the detector
(different isotopes of Al and O 
in the {CSR} detector \cite{CRESST:2017ues}), 
$N_{A}$ is the number of nucleus $A$ per unit mass, 
$\sigma_{\chi A}$ is interaction cross section between DM 
and nucleus $A$, 
as given in Eq.\ (\ref{eq:DMxsec}), 
and $f(v)$ is the DM velocity distribution, 
as given in Eq.\ \eqref{Eq:v-distribution}. 
For the SI case, we do not need to distinguish different 
isotopes. For the SD case, we only need to consider the isotopes 
with non-zero spin: $^{17}$O and $^{27}$Al. 
Since the natural abundance of $^{17}$O is 0.04\% \cite{lodders1998planetary}, the dominant contribution to SD scatterings comes from $^{27}$Al, whose natural abundance is 100\%.
The number of the isotope are obtained by multiplying 
the number of the element with the natural abundance.

We obtain the CSR experimental data from the 
data file provided at the website 
\cite{CSR-data}. 
We bin the data in the energy range of $[19.7, 600]$ eV 
with a bin width of 5 eV, following Ref.\ \cite{CRESST:2017ues}.
We further carry out a $\chi^2$ analysis to calculate the lower boundary of the CSR exclusion region, 
where $\chi^2$ is obtained via the same method as in 
the XQC case, namely computed via Eq.\ \eqref{Eq:chi2} 
with the quantities substituted with the CSR 
predictions and data such that 
$U_i$ is the number of events in the CSR data \cite{CRESST:2017ues}. 
We exclude the data bins with null results 
in the total $\chi^2$ analysis; 
we find that such a method 
leads to consistent results with
Ref.\ \cite{CRESST:2017ues}.
The {90\%} CL constraints on the 
lower boundary of the exclusion region 
are then obtained by 
requiring a total $\chi^2=2.71$.

\subsection{The upper boundary of the CSR exclusion region}

We adopt ``method-a'' in Ref.\ \cite{Emken:2018run} to compute the upper boundary of the CSR exclusion region. 
This method demands that all DM lose energy in the atmosphere 
so that after traversing the atmosphere, the maximum 
recoil energy of DM is below the CSR detection 
threshold ($E_R^{\rm min}=19.7$ eV).
This method leads to a satisfactory limit as compared 
with the MC method as shown in figure 5 of 
Ref.\ \cite{Emken:2018run}.

In our analysis, we consider the DM particle with 
the largest possible initial velocity before 
entering the atmosphere, 
$v_\chi^i = v_{\rm esc} +  v_\odot \simeq 800\, {\rm km/s}$, 
where $v_\odot\simeq 220$ km/s is the velocity of the sun. 
We then obtain the final kinetic energy $E_\chi^f$ of DM by 
computing the total energy loss
as the DM particle traverses the atmosphere 
via the differential 
equation given in Eq.\ \eqref{Eq:dedx}. 
In our analysis, we adopt the same atmosphere model as 
in Ref.\ \cite{Mahdawi:2018euy}; 
a brief description of the atmosphere model is given 
in appendix \ref{sec:atmosphere-model}.
We obtain the upper limit by
demanding the maximum recoil energy of DM, 
$E_R^{\rm max}=4 E_\chi^f m_\chi m_N/(m_\chi + m_N)^2$, 
to be below the CSR detection 
threshold, $E_R^{\rm min}=19.7$ eV.

\section{Comparison of XQC and {CSR} limits to other experimental limits}
\label{sec:constraint-mass-fraction}

We compare the XQC and CSR limits {computed} in our analysis 
with other 
experimental constraints, 
for DM models with different DM mass fractions and for both the SI and SD cases. 
These include:   
(1) the balloon experiment: RRS \cite{Rich:1987st}; 
(2) experiments at  the surface of the Earth: EDELWEISS \cite{EDELWEISS:2019vjv}; 
(3) underground experiments near the surface of the Earth: CDMS surface \cite{CDMS:2000lgz} and DAMIC shallow site run \cite{DAMIC:2016lrs};
(4) deep underground experiments considered in Ref.\ \cite{Hooper:2018bfw}; 
(5) CMB {and} Lyman-$\alpha$ constraints 
\cite{Boehm:2001hm,Cyburt:2002uw,Chen:2002yh,Boehm:2004th,Dvorkin:2013cea,Gluscevic:2017ywp,Boddy:2018wzy,Xu:2018efh,Buen-Abad:2021mvc,Boddy:2022tyt,Hooper:2022byl}; 
(6) boosted DM constraints 
\cite{Bringmann:2018cvk,
Cappiello:2018hsu,
Kim:2019had,
Cappiello:2019qsw,
Dent:2019krz,
Krnjaic:2019dzc,
Wang:2019jtk,
Guo:2020drq,
Ge:2020yuf,
Cao:2020bwd,
Xia:2020apm,
Flambaum:2020xxo,
PROSPECT:2021awi,
Bell:2021xff,
Feng:2021hyz,
Wang:2021nbf,
Xia:2021vbz,
Wang:2021jic,
PandaX-II:2021kai,
CDEX:2022fig,
Granelli:2022ysi,
Xia:2022tid}.

\begin{figure}[tbp]
    \centering
    \includegraphics[width=0.9\textwidth]{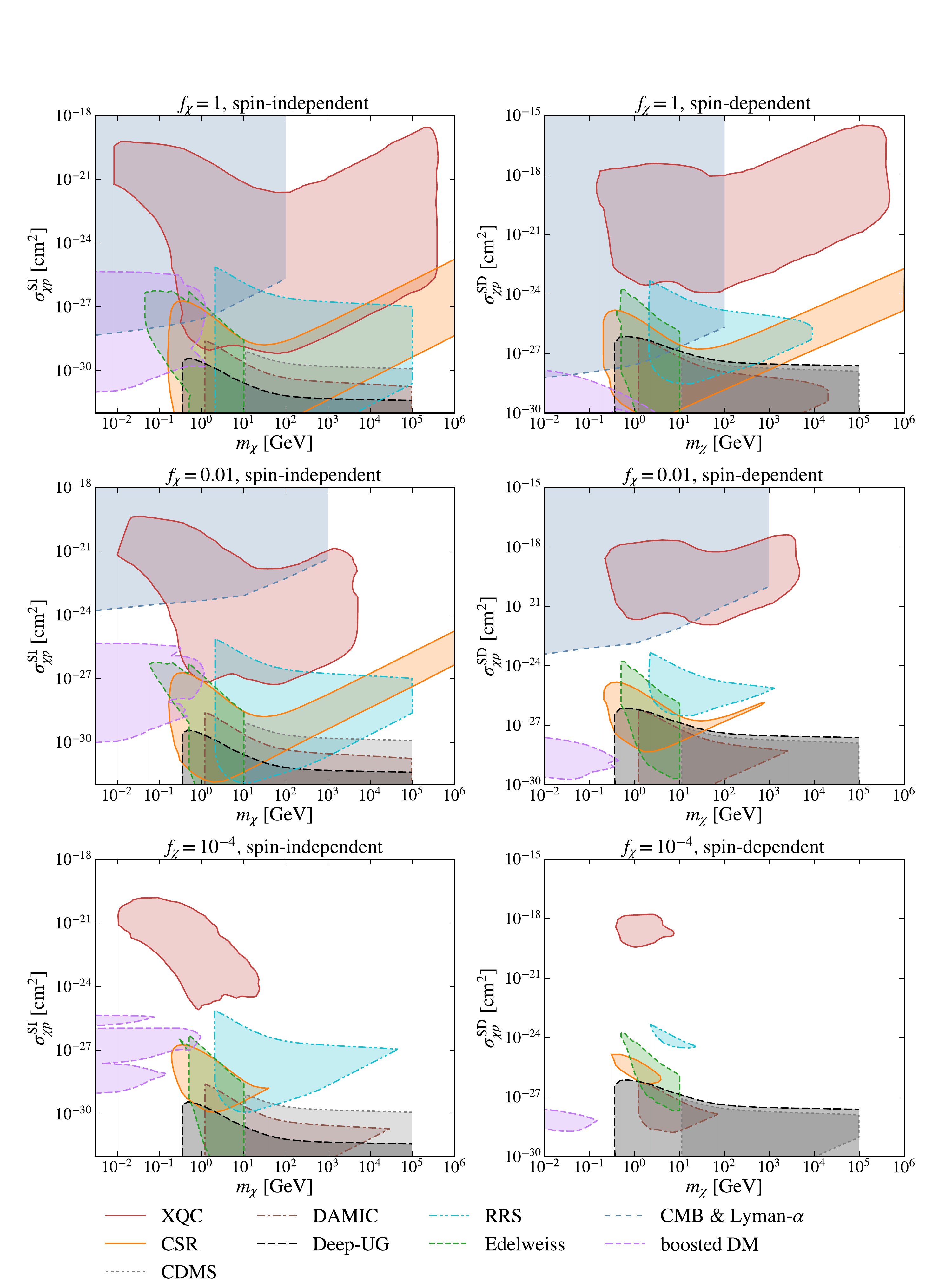}
    \caption{Experimental exclusion regions for velocity-independent 
    DM-proton cross sections, 
    both in the SI case ({\it left}) 
    and in the SD ({\it right}) cases, 
    with different mass fractions: 
    $f_\chi = 1$ ({\it top}), 
    $f_\chi = 10^{-2}$ ({\it middle}), and  
    $f_\chi = 10^{-4}$ ({\it bottom}). 
    The exclusion region of XQC (our results) and CSR (our results) are shown as the red and orange regions respectively. Other exclusion regions include RRS (cyan), CDMS (gray), DAMIC (brown) and deep-underground experiments (black) \cite{Hooper:2018bfw}, EDELWEISS (green) \cite{EDELWEISS:2019vjv},
    and CR-boosted DM (purple) \cite{Bringmann:2018cvk,Cappiello:2019qsw,PROSPECT:2021awi,Wang:2021nbf,PandaX-II:2021kai,CDEX:2022fig}. 
    The CMB and Lyman-$\alpha$ limits are adopted
    from Lyman-$\alpha$ (top) \cite{Buen-Abad:2021mvc},
    and from CMB (middle) \cite{Boddy:2022tyt}.
    }
    \label{fig:exclusion}
\end{figure}

Fig.\ (\ref{fig:exclusion}) shows the various experimental constraints 
on velocity-independent DM-proton cross sections, 
for both SI and SD interactions, 
and for three different mass fractions: 
$f_\chi = 1$, $f_\chi = 10^{-2}$, 
and $f_\chi = 10^{-4}$. 
In the $f_\chi = 1$ case, 
we adopt the EDELWEISS exclusion region from Ref.\ \cite{EDELWEISS:2019vjv}, 
the boosted DM constraints from Ref.\ \cite{Bringmann:2018cvk,Cappiello:2019qsw,PROSPECT:2021awi,Wang:2021nbf,PandaX-II:2021kai,CDEX:2022fig}, 
and constraints of other experiments from Ref.\ \cite{Hooper:2018bfw}.
For the boosted DM constraints in the SI case,  
we consider limits from KamLAND \cite{Cappiello:2019qsw}, PROSPECT \cite{PROSPECT:2021awi} , PANDAX-II \cite{PandaX-II:2021kai}, CDEX \cite{CDEX:2022fig}, XENON1T and MiniBooNE \cite{Bringmann:2018cvk};
For the boosted DM constraints in the SD case,  
we consider limits from 
Borexino \cite{Bringmann:2018cvk} and XENON1T \cite{Wang:2021nbf}.\footnote{The boosted DM constraints are typically 
significant for light DM, 
where cosmological/astrophysical constraints 
can be also important; see e.g., Ref.\ \cite{Elor:2021swj}
for a recent study on the maximally allowed 
cross section of light DM and 
to reach such values in a viable particle model.}
We rescale the above limits given in the $f_\chi =1$ case to the $f_\chi =10^{-2}$ and $f_\chi =10^{-4}$ cases in Fig.\ (\ref{fig:exclusion}).\footnote{The CMB and Lyman-$\alpha$ constraints have non-trivial dependencies on the DM mass fraction, which we discuss separately.}
For mass fractions less than $f_\chi =1$ we use 
the following approximation to compute the exclusion regions: 
The upper boundary of the exclusion region is assumed to be the same as the $f_\chi =1 $ case; 
the lower boundary of the exclusion region is computed by rescaling the lower boundary in the 
$f_\chi = 1$ case by the mass fraction such that 
\begin{equation}
    \sigma_l (f_\chi) = {\sigma_l (f_\chi = 1) \over f_\chi^k}, 
    \label{Eq:lower-rescale}
\end{equation}
where $\sigma_l (f_\chi =1 )$ is the lower boundary on the cross section in the $f_\chi =1 $ case, 
and $k=1(1/2)$ for DM with expected velocity in the DM halo (for the boosted DM). 
The rescaling for the halo DM for 
most cases has been confirmed to be a fairly reasonable 
approximation in our XQC simulations; 
the rescaling for the 
boosted DM is due to the fact 
that the event rate of the boosted DM is proportional to 
$\rho_\chi \sigma_{\chi p}^2$ \cite{Wang:2021jic}.

For CMB and Lyman-$\alpha$ constraints shown in Fig.\ (\ref{fig:exclusion}),
we adopt the Lyman-$\alpha$ constraints from Ref.\ \cite{Buen-Abad:2021mvc}
and the CMB limits from Ref.\ \cite{Boddy:2022tyt} and with further
rescalings.
In the $f_\chi = 1$ case, 
because the Lyman-$\alpha$ limits are
stronger than the CMB limits,
we only show the Lyman-$\alpha$ constraints 
from Ref.\ \cite{Buen-Abad:2021mvc} for
both SI and SD cases.

In the $f_\chi = 0.01$ case,
the CMB limits become stronger 
than the Lyman-$\alpha$ constraints, 
since the Lyman-$\alpha$ constraints are expected to 
decrease with the DM mass fraction much faster than  
the CMB limits.\footnote{The different dependencies on the 
DM mass fraction can be seen in Refs.\ \cite{Boddy:2018wzy,Hooper:2022byl}. 
The CMB limits  
are expected to decrease inverse-linearly with 
the DM mass fraction. 
Ref.\ \cite{Boddy:2018wzy} pointed out that 
such a re-scaling should hold for $f_\chi \gtrsim 2\%$ 
in the $n=-4$ case. 
Recently, Ref.\ \cite{Hooper:2022byl} has provided 
for the $m_\chi=1$ GeV case that
the CMB+BAO (Lyman-$\alpha$) limits can be rescaled 
with $\sim f_\chi^{-1}$ ($\sim f_\chi^{-3}$) in the range of $0.01\leq f_\chi\leq 1$.} 
Thus we obtain the CMB constraints in the $f_\chi=0.01$ case 
by inverse-linearly rescaling the limits analyzed in the $f_\chi=1$ case
from Ref.\ \cite{Boddy:2022tyt} for both SI and SD cases. 
We note that recently Refs.\ \cite{Buen-Abad:2021mvc,Hooper:2022byl}
have combined the CMB constraints with the 
baryon acoustic oscillations (BAO) limits, 
leading to a better upper bound on the cross section 
than the CMB only, by a factor 
of $\lesssim 2.5$ ($\sim 1.5$) for  
light (heavy) DM mass.

Although we do not show the CMB constraints for even 
smaller DM mass fractions, 
we note that the stringent CMB constraints on 
DM of the $n=-4$ case  (e.g., millicharged DM)
disappear if the DM mass fraction is less 
than $0.2\%$, 
since such a small DM component is distinguishable 
from baryons {\cite{Boddy:2018wzy}}.
Thus we expect the CMB and Lyman-$\alpha$ constraints
to be further weakened as the DM mass fraction decreases
to even smaller values, 
or even to disappear.

In the SI case with $f_\chi =1$, DM with 
cross section shown in Fig.\ (\ref{fig:exclusion}) 
and with mass in the range of  
$[10, 10^5]$ GeV are ruled out by combining all experimental 
constraints. 
However, in the SD case with $f_\chi = 1$, 
there are two parameter regions as shown in 
Fig.\ (\ref{fig:exclusion}) 
are still allowed. 
The allowed parameter space increases significantly 
as the mass fraction $f_\chi$ is changed from unity 
to $10^{-2}$ and further to $10^{-4}$. 
In the SD case with $f_\chi = 10^{-4}$, 
almost the entire parameter space with 
$m_\chi \gtrsim 10$ GeV and 
$\sigma_{\chi p} \gtrsim 10^{-27}$ cm$^2$ 
seems to be allowed.

\section{Summary}
\label{sec:summary}

In this paper, we use MC simulations to calculate the XQC exclusion regions 
on SIDM 
with different mass fractions, 
with different velocity-dependent cross sections, 
and with both SI and SD interactions. 
It is found that, 
to a good approximation, 
there is a (inverse) linear relation 
between the lower/right boundary of the XQC exclusion region 
and the mass fraction. 
For the upper/left boundary of the XQC exclusion region, 
the MC simulations reveal some 
non-trivial dependence on the mass fraction. 
We note that our results agree with
previous analyses that used MC simulations 
to compute the XQC limits, such as 
Refs.\ \cite{Erickcek:2007jv,Mahdawi:2017cxz,Mahdawi:2018euy}, 
when the same assumptions on the DM models and on 
the response of the XQC detector are made. 
The primary goal of our analysis is to 
study the entire XQC excluded parameter space 
for 
different types of DM interaction cross sections and for different DM
mass fractions, which have not been 
fully explored in previous studies. 
To facilitate the intensive MC simulations in 
modeling the interactions of DM with the XQC detector 
for the full parameter space, 
we have neglected 
the fact that the aluminum shield can be somewhat transparent 
in a small portion of the parameter space at low DM mass, 
and the thermalization efficiency of the silicon detector, 
which, however, has not been experimentally measured yet.

For the velocity-dependent cross sections, 
we consider four different cases: $n$ = \{0, 2, -2, -4\}
in the form of $\sigma_{\chi p} = \sigma_{\chi p}^0 v^n$. 
We find that most of the exclusion regions for different $n$'s 
overlap with each other. 
However, there are also some differences among them: 
For the $n=-4$ case, the lower boundary of the XQC exclusion region  
is a little lower than other cases; 
for the $n=2$ case, the upper boundary for heavy mass 
is much higher than 
other cases.
Also, for the $n=2$ case, 
there exists a large portion of parameter space 
deep inside the XQC exclusion region that 
is unconstrained by XQC;
this is due to the unused energy range in the XQC data 
and the ``focusing'' effects of the filters on the DM velocity 
so that the DM velocity is localized in a narrow range.

{We compare the XQC exclusion regions in the SI and the SD cases 
and find that the XQC exclusion regions in the SD case are quite different 
from the SI case. 
Although the lower boundary of the XQC exclusion region in the SD case 
is similar to the SI case in the shape, it is 
larger by a factor of $10^5$. 
For the upper boundary, 
the XQC exclusion region in the SD case is very different from the 
SI case and there is no simple re-scaling relation between them. 
We also find that 
the XQC exclusion regions disappear 
if the mass fraction is significantly small, 
$f_\chi \lesssim10^{-5} (10^{-4})$ for the SI (SD) case.

We also analyze the CSR constraints for different DM mass fractions
both for the SI and for the SD cases. 
We further compare the XQC and CSR limits computed in our analysis, 
with various other experimental constraints. 
We find that the allowed parameter space 
increases when the mass fraction decreases; 
for a sufficiently small mass fraction, such as 
$f \lesssim 10^{-4}$, 
a very large portion of parameter space is allowed.

\acknowledgments

The work was supported in part by the 
National Natural Science Foundation of China under Grant Nos.\ 
12275128, 
11775109, 
and 12147103, and by the Fundamental Research Funds for the Central Universities.

\appendix
\renewcommand{\appendixname}{Appendix~\Alph{section}}

\section{Nucleus with spin}
\label{sec:natural-abundance}

Only nuclei with spin participate in the SD interactions. 
Table \ref{tab:spin-imformation} shows the nuclei with spin 
that are used in our analysis.

\begin{table}[tbp]
    \centering
    \begin{tabular}{|c|c|c|c|c|}
    \hline
     Element  &  Abundance & $J_A$ &$\left\langle\mathbf{S}_{p}\right\rangle$ &  $\left\langle\mathbf{S}_{n}\right\rangle$  \\ \hline\hline
     $^{27}$Al   &  100\% &5/2 & 0.343 & 0.0296  \\ \hline
     $^{13}$C   &  1.1\% & 1/2 & 0 & -0.167 \\ \hline
     $^{29}$Si   & 4.7\% &1/2 & -0.002  & 0.13 \\ \hline
     $^{17}$O   & 0.04\% &5/2 & 0 & 0.5 \\ \hline
     $^{14}$N   & 99.6\% & 1 & 0.5 & 0.5 \\ \hline
    \end{tabular}
    \caption{Nuclei that are considered in the SD calculations 
    in the analysis. 
    The natural abundance of the nuclei is obtained in Ref.\ \cite{lodders1998planetary}, and the spin information ($J_A$, $\left\langle\mathbf{S}_{p}\right\rangle$, $\left\langle\mathbf{S}_{n}\right\rangle$) is obtained from Refs.\ \cite{Bednyakov:2004xq,Hooper:2018bfw}. We have neglected $^{15}$N, since its contribution is too small compared with $^{14}$N.}
    \label{tab:spin-imformation}
\end{table}

\section{DM velocity after scattering}
\label{sec:scattering:angle}

The final velocity $\bm v_f$ of the DM particle in the lab frame 
after a single scattering with the target nucleus $N$ is given by 
\be
{\bm v_f} = {\bm v_f^{\rm *}} + {\bm u}, 
\ee
where ${ \bm v_f^{\rm *} }$ is the DM final velocity in the CM frame, 
and 
${\bm u}$ is the velocity of the CM frame in the lab frame, 
which is given by 
\be 
\bm u = \frac{m_\chi v}{m_\chi+m_N}\left(\begin{matrix}
                                \sin\theta\cos\phi \\
                                \sin\theta\sin\phi \\
                                \cos\theta
                                \end{matrix}\right),
\ee 
where $m_\chi$ is the DM mass, 
$m_N$ is the target nuclear mass, 
and $v$, $\theta$ and $\phi$ 
are the magnitude, polar angle and 
azimuthal angle of the DM initial velocity in the lab frame respectively.
Setting the $z$ axis of the CM frame along the initial velocity of the DM particle, 
one has 
\be    
{ \bm v_f^{\rm *} } = \frac{m_N v}{m_\chi+m_N}
            \left(\begin{matrix}\cos\phi(\sin\alpha\cos\beta\cos\theta+
                            \sin\theta\cos\alpha) - \sin\phi\sin\alpha\sin\beta \\
                            \sin\phi(\sin\alpha\cos\beta\cos\theta+
                            \sin\theta\cos\alpha) +\cos\phi\sin\alpha\sin\beta  \\
                            \cos\theta\cos\alpha - \sin\theta\sin\alpha\cos\beta
                        \end{matrix}\right),
\ee
where $\alpha$ and $\beta$ are the polar angle and the azimuthal angle  
of the final DM velocity in the CM frame respectively.  
The {percentage} energy loss of the DM particle during 
the scattering process is then given by  
\be
    {{\Delta E \over E_\chi^i}  
    \equiv {E_\chi^i - E_\chi^f \over E_\chi^i} }
    = \frac{2m_\chi m_N}{(m_\chi + m _ N)^2}(1 - \cos\alpha), 
    \label{Eq:energytransfer}
\ee
where {$E_\chi^i$ ($E_\chi^f$)} the initial (final) kinetic energy of the 
DM particle in the lab frame.

\section{XQC data}
\label{sec:XQC-data}

The XQC data 
are binned into thirteen different energy 
bins in Ref.~\cite{Erickcek:2007jv}; 
see table 1 of Ref.~\cite{Erickcek:2007jv}
for the energy ranges and the number of counts 
for each bin.
The detection efficiency is small for the low energy 
measurements. Following Ref.\ \cite{Erickcek:2007jv}, 
we use efficiency $f = 0.3815 \, (0.5083)$ for the 
first (second) energy bin, and $f=1$ for the eleven 
high energy bins.

\section{The number of DM particles inside the FoV of XQC}
\label{sec:Nchi}

The total number of DM particles
(incident on the surface area of the detector 
or of the filter) 
in the XQC FoV 
during the data-takings is given by 
\be
N_\chi =f_\chi {\rho_\chi\over m_\chi} S t \int f_B(|\bm{v'}+\bm{v}_{\rm det}|)\ v'\cos\theta\, v'^2d v' d\Omega, 
\label{Eq:Nchi-estimate1}
\ee
where 
$\rho_\chi=0.3$ GeV/cm$^3$ 
is the local DM mass density, 
$m_\chi$ is the DM mass, 
$f_\chi$ is the mass fraction, 
$S$ is the surface area of the detector or 
the filter,
$t=100.7$ sec is the detection time, 
$\bm{v'}$ is the DM velocity in the lab frame,
$\bm{v}_{\rm det}=(39.14,230.5,3.573)$ km/s \cite{Erickcek:2007jv} 
is the velocity of the detector 
in the halo frame in he galactic coordinate system,
and $f_B$ is the truncated Boltzmann distribution of 
the DM velocity in the halo frame 
as given in Eq.\ \eqref{Eq:v-distribution}. 
In our analysis, we take 
$S=0.34$ $\rm{cm^2}$ ($S=7.1$ $\rm{cm^2})$ for the detector (filter) area \cite{McCammon:2002gb}. 
In the coordinate system where the z-axis points to the 
opposite direction of the center of the FoV, 
the integral is performed in the range of 
$0<\theta<\cos^{-1}(1-(2\pi)^{-1})$ and $0<\varphi<2\pi$.
Thus, we have 
\be
    N_\chi = 3.52\times 10^7f_\chi \, \frac{\rm GeV}{m_\chi}\frac{S}{0.34\, {\rm cm}^2}.
\label{Eq:Nchi-estimate2}
\ee

\section{DM energy loss through an overburden}
\label{sec:energy-loss}

In this section, we analyze the energy loss of the DM 
when traversing an overburden.
See also Refs.\
\cite{Emken:2018run} 
\cite{Mahdawi:2018euy} 
for similar analysis. 
The energy loss of DM, 
when passing through an overburden, 
can be computed via 
\begin{equation}
\frac{d E_{\chi}}{d x}=-\sum_{N} n_N \int_{0}^{E_{R}^{\max }} d E_{R} E_{R} \frac{d \sigma_{\chi N}}{d E_{R}},
\label{Eq:dedx}
\end{equation}
where 
$E_\chi$ is the DM kinetic energy, 
$x$ is the distance traveled by DM, 
$n_N$ is the number density of nucleus $N$, 
$E_R$ is the nuclear recoil energy, 
${d \sigma_{\chi N}}/{d E_{R}}$ is the differential cross section, 
and $E_{R}^{\max }= {2\mu_{\chi N}^{2} v_\chi^{2}/m_{N}}$ 
is the maximum recoil energy where 
$v_\chi$ is the DM velocity,  
$m_N$ is the nucleus mass, 
and $\mu_{\chi N} = m_\chi m_N/(m_\chi + m_N)$ 
is the reduced mass with $m_\chi$ being the DM mass. 
Here the summation runs over different nuclei in the 
overburden. 
For a velocity-dependent cross section that is proportional 
to $v^n$, 
one has 
\begin{equation}
\frac{d \sigma_{\chi N}}{d E_{R}}=\frac{\sigma_{\chi N}^{\mathrm{tot}} }{E_{R}^{\max}} 
= \frac{\sigma_{\chi N}^{0} v_\chi^n }{E_{R}^{\max}}, 
\label{Eq:dsigmadER}
\end{equation}
where 
$\sigma_{\chi N}^{\mathrm{tot}}$ is the 
total DM-nucleus cross section, 
and we have assumed that the scattering cross section is isotropic in the 
CM frame. 
Substituting Eq.\ \eqref{Eq:dsigmadER} into Eq.\ (\ref{Eq:dedx}) 
and performing the integral on $E_R$, one has 
\begin{equation}
\begin{aligned}
\frac{d E_{\chi}}{d x} 
= - E_T^2 \left( \frac{2E_\chi}{m_\chi}\right)^{\frac{n}{2}+1},  
\end{aligned}
\label{Eq:energy-loss-v}
\end{equation}
where we have used $E_\chi = m_\chi v_\chi^2/2$ and defined the following quantity 
\begin{equation}
    E_T^2 \equiv \sum_{N} \frac{ n_{N}  \mu_{\chi N}^{2} \sigma_{\chi N}^{0}}{m_{N}}. 
\end{equation}
The final energy of DM after passing through an overburden 
can then obtained by 
solving Eq.\ \eqref{Eq:energy-loss-v} 
\begin{equation}
E^f_{\chi}= E_\chi^i \left\{
\begin{aligned}
&\left[ (v_\chi^i)^n
\frac{n d E_T^2}{m_\chi}
 + 1 \right]^{-\frac{2}{n}}, 
\quad & n\neq0, \\
 & \exp\left[ -2d E_T^2/m_\chi
 \right], \quad & n=0 \\
\end{aligned}\right.
\label{Eq:energy-loss}
\end{equation}
where $d$ is the distance traveled by DM in the overburden, 
$v_\chi^i$ is the initial DM velocity, and 
$E^i_{\chi}$ ($E^f_{\chi}$) is the 
initial (final) kinetic energy of DM.

\subsection{The upper boundary of the XQC exclusion region}
\label{sec:analy-upper}

We can use ``method-a'' in {Ref.\ \cite{Emken:2018run}} 
to estimate 
the upper boundary of the XQC exclusion region: 
We demand that DM particles with the largest possible velocity, 
$v_\chi^i = v_{\rm esc} +  v_\odot \simeq 800\, {\rm km/s}$, 
lose energy through the 5 filters such that the 
final energy after those filters are below the 
energy threshold of the XQC experiment, 
which is 29 eV. 
We compare such an estimate with our MC simulations 
in Fig.\ (\ref{fig:XQC-upper-compare}). 
We find that the estimated method 
is consist with our MC simulations 
for DM masses larger than $\sim 1000$ GeV. 
For small DM masses, however, 
the estimates are a little 
larger than results in our MC simulations.

\begin{figure}[tbp]
    \centering
    \includegraphics[width=0.45\textwidth]{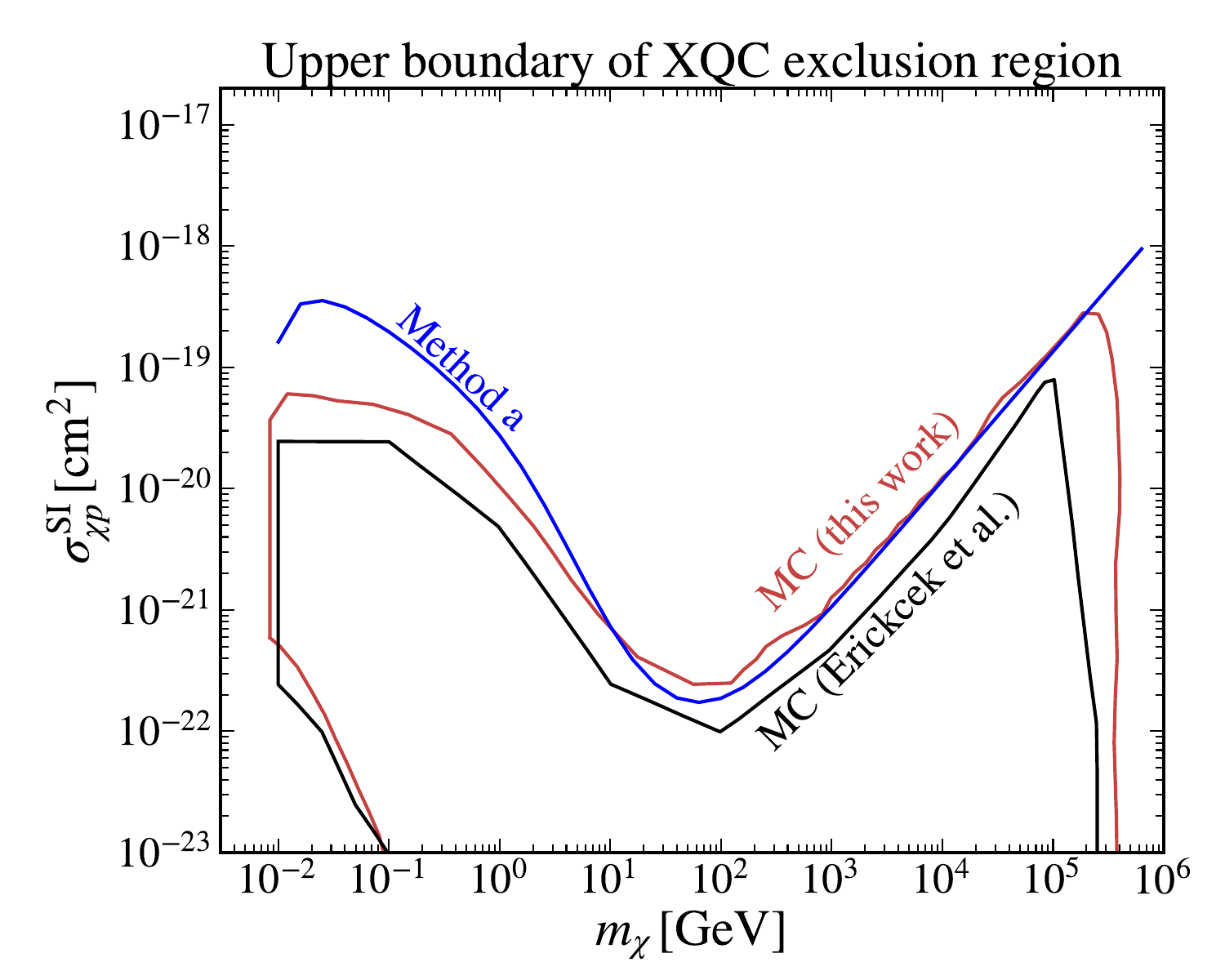}
    \caption{Comparison of different methods of computing the 
    upper boundary of the XQC exclusion region 
    in the SI and $n=0$ case: 
    our MC simulation (red), 
    the analytical ``method-a'' adopted from Ref.\ \cite{Emken:2018run} 
    (blue), 
    and the 90\% CL exclusion region given by Ref.\ \cite{Erickcek:2007jv} 
    (black). 
    }
    \label{fig:XQC-upper-compare}
\end{figure}

\subsection{The allowed band in the $n=2$ case}
\label{sec:n2-band}

\begin{figure}[tbp]
    \centering
    \includegraphics[width=0.45\textwidth]{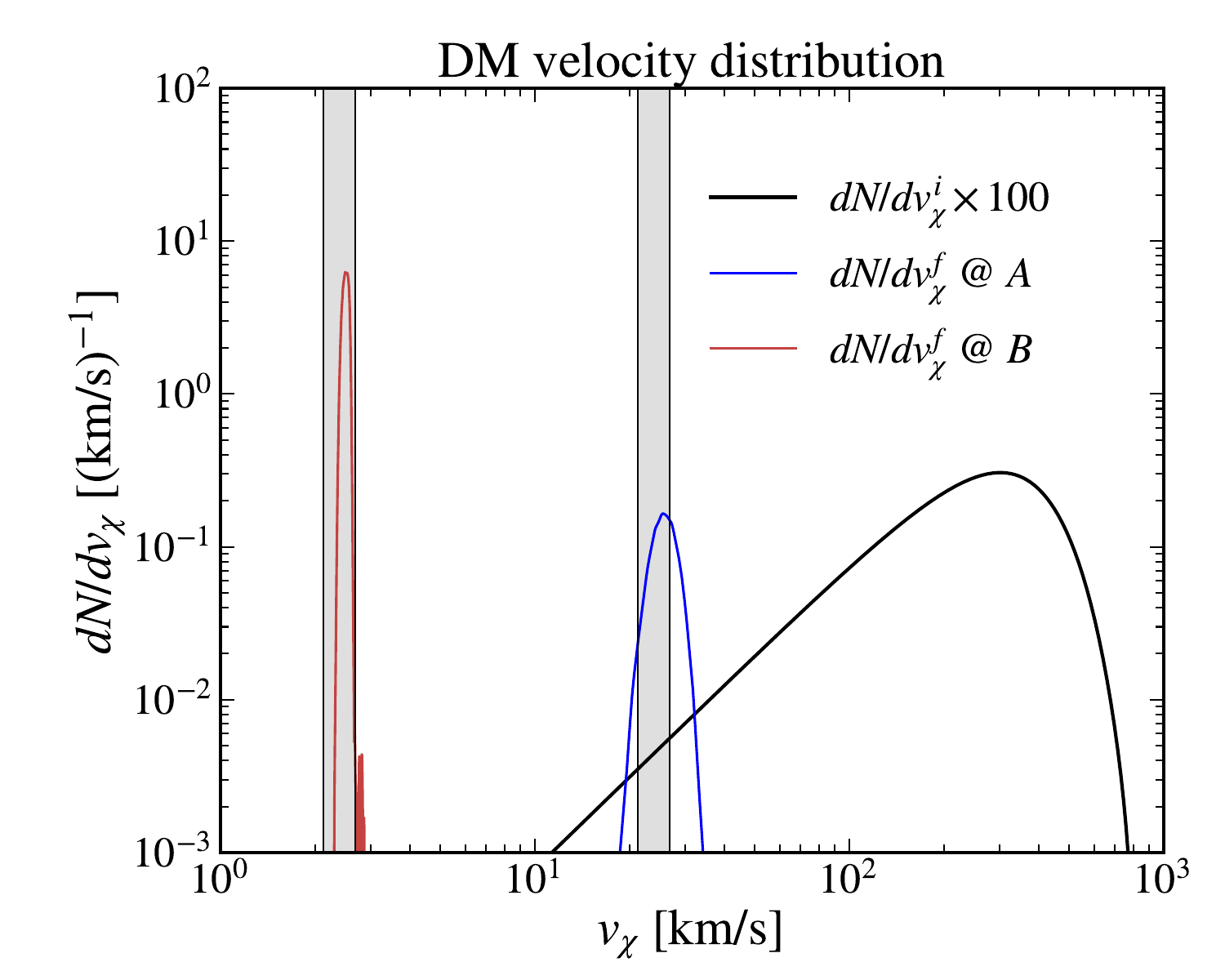}
    \includegraphics[width=0.45\textwidth]{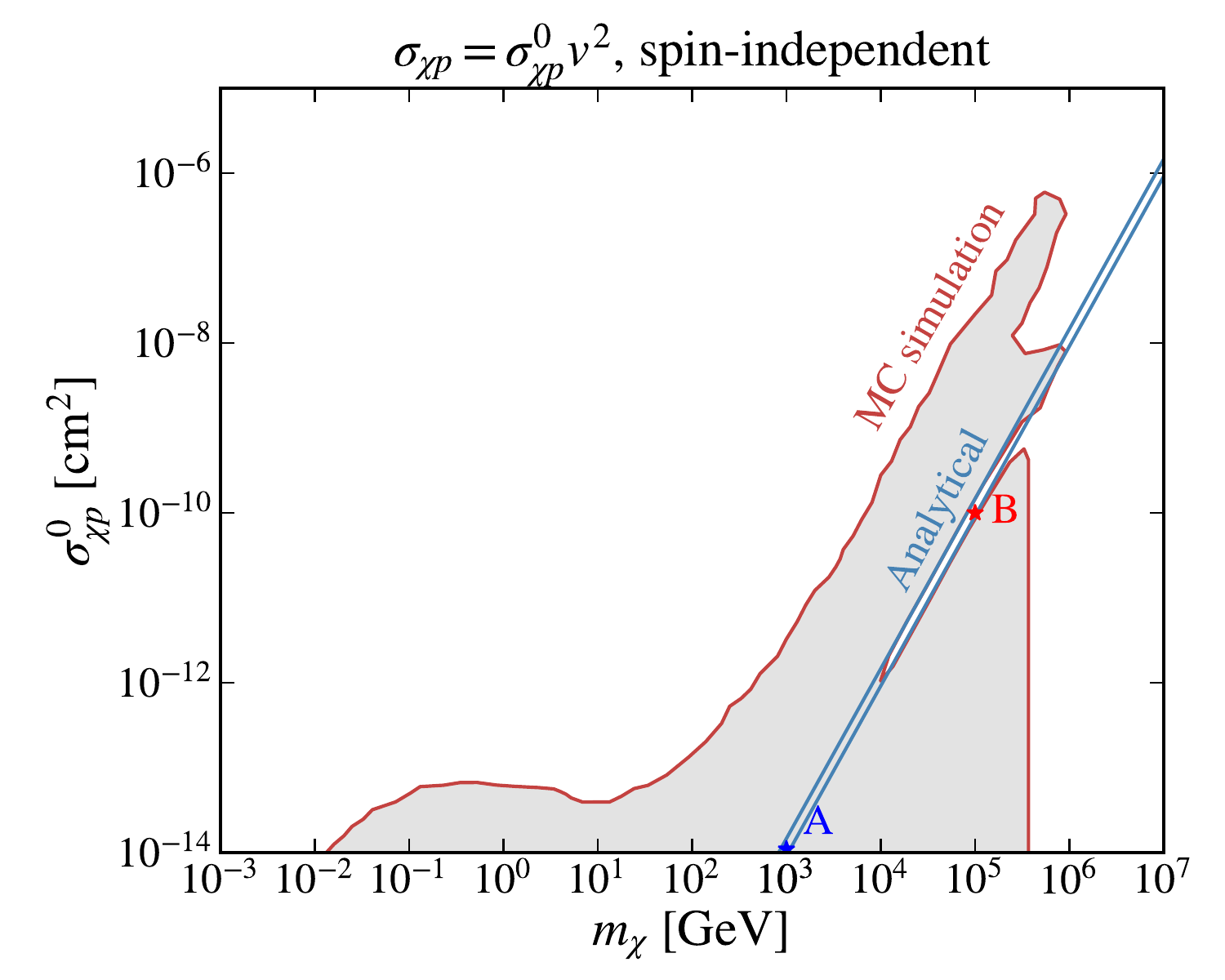}
    \caption{{\bf Left panel:} 
    The DM velocity distributions in the lab frame for 
    DM before penetrating the XQC filters (black) 
    and for two DM model points after penetrating 
    the XQC filters: model $A$ (blue) 
    and model $B$ (red). 
    The vertical gray bands indicate the velocity range 
    where the kinetic energy of the DM is in 
    the range of $[2505,4000]$ eV. 
    {\bf Right panel:} The XQC exclusion region 
    in the SI case (red boundary) 
    in the parameter space spanned by 
    $\sigma_{\chi p}^0$ and the DM mass in the $n=2$ case. 
    The two blue lines are 
    obtained by equating $E_c$ in Eq.\ \eqref{Eq:Ec-evaluation} 
    to 2505 eV and 4000 eV respectively. 
    The benchmark model point $A$ with 
$(m_\chi,\sigma^0_{\chi p})= (10^3\, \rm{GeV},10^{-14}\, \rm{cm^2})$ 
and the model point $B$ with 
$(m_\chi,\sigma^0_{\chi p}) = (10^5\, \rm{GeV},10^{-10}\, \rm{cm^2})$ 
are shown as stars.} 
    \label{fig:energy-spectrum}
\end{figure}

As shown in Fig.\ (\ref{fig:XQC-mf-SI}) and Fig.\ (\ref{fig:XQC-mf-SD}), 
for the $n=2$ case, 
there exist parameter space, as described by Eq.\ \eqref{eq:band}, 
where the XQC constraints are ``mysteriously'' gone.  
This can be explained by the energy loss calculations presented here 
and the XQC data bins. 
In the $n=2$ case, Eq.\ \eqref{Eq:energy-loss} becomes 
\be
    \frac{1}{E_\chi^{f}} = \frac{1}{E_c} + \frac{1}{E_\chi^{i}},
    \label{Eq:energy-loss-v2}
\ee
where $E_c = m_\chi^2/(4 d E_T^2)$.
Thus, for XQC, one has {(for $m_\chi\gg m_N$)}
\begin{equation}
E_c \simeq 3.67\times 10^3 (1.17\times 10^7) \,{\rm eV} \, \left[\frac{m_\chi}{10^4 \, \rm GeV}\right]^2 \,
\frac{10^{-12}\, \rm cm^2}{\sigma_{\chi p}^0}, 
\label{Eq:Ec-evaluation}
\end{equation}
for the SI (SD) case. For $E_\chi^{i} \gg E_c$, 
the DM final energy becomes 
localized in the vicinity of $E_c$, namely $E_\chi^{f} \approx E_c$, 
which is independent of the DM initial energy $E_\chi^i$.
Under such circumstances, if $E_c$ (obtained by summing the 5 filters) 
falls inside $[2505,4000]$ eV 
(the energy range discarded by XQC \cite{Erickcek:2007jv}), these DM particles do not give rise 
to detectable events in XQC. 
Thus, by equating $E_c$ in Eq.\ \eqref{Eq:Ec-evaluation} 
to 2505 eV and 4000 eV, 
we obtain two curves in the parameter space 
spanned by $m_\chi$ and $\sigma_{\chi p}^0$, 
which are shown in the right panel of 
Fig.\ (\ref{fig:energy-spectrum}) for the SI case; 
the two lines obtained via Eq.\ \eqref{Eq:Ec-evaluation} 
accurately describe the MC results in the mass range of 
$10^4-3\times 10^5$ GeV.

To understand why the analytical results via Eq.\ \eqref{Eq:Ec-evaluation} 
fail below $\sim 10^4$ GeV, 
we select two benchmark model points on the right panel figure of 
Fig.\ (\ref{fig:energy-spectrum}): 
model point $A$ with 
$(m_\chi,\sigma^0_{\chi p})= (10^3\, \rm{GeV},10^{-14}\, \rm{cm^2})$ 
and model point $B$ with  
$(m_\chi,\sigma^0_{\chi p}) = (10^5\, \rm{GeV},10^{-10}\, \rm{cm^2})$, 
and compare their velocity distributions after passing the 
five XQC filters on the left panel figure of 
Fig.\ (\ref{fig:energy-spectrum}). 
The kinetic energy of DM particles 
in the model point B 
has a very narrow distribution so that almost
all the DM particles are inside 
the energy range of $[2505,4000]$ eV: 
252 DM particles inside the 
$[2505,4000]$ eV range with less than 1 event 
outside during the XQC data-taking.
However, the kinetic energy in the 
model point A has a rather extended distribution 
so that a significant fraction of DM particles 
are outside the energy range of 
$[2505,4000]$ eV: 
14553 and 8807 events are inside 
and outside the  
$[2505,4000]$ eV range during the 
XQC data-taking respectively.  
Thus the DM particles in the model point A 
are no longer all inside 
the XQC discarded data bin of 
$[2505,4000]$ eV, 
which results in the failure of the simple 
analytic method.

\section{Atmosphere model}
\label{sec:atmosphere-model}

We use the same atmosphere model as in Ref.\ \cite{Mahdawi:2018euy}. 
For completeness, we provide a brief description 
of the atmosphere model in this appendix. 
The atmosphere from the Earth surface to the altitude of $z=84.8$ km 
is divided into 7 layers \cite{Mahdawi:2018euy},
as shown in Table \ref{tab:atmosphere}. 

\begin{table}[tbp]
\centering
\begin{tabular}{|c|c|c|c|}
\hline layer & $z_{\rm min}$ (km)  & $z_{\rm max}$ (km)  & $\rho_{\rm atm}(z)/\rho_{\rm atm}(z=0)$ \\
\hline 1 & 0 & 11 & $(1-z / 44.31)^{4.256}$ \\
\hline 2 & 11 & 20 & $0.297 \exp ((11-z) / 6.34)$ \\
\hline 3 & 20 & 32 & $(0.978+z / 201.02)^{-35.16}$ \\
\hline 4 & 32 & 47 & $(0.857+z / 57.94)^{-13.2}$ \\
\hline 5 & 47 & 51 & $1.165 \times 10^{-3} \exp ((47-z) / 7.92)$ \\
\hline 6 & 51 & 71 & $(0.8-z / 184.8)^{11.2}$ \\
\hline 7 & 71 & 84.8 &$(0.9-z / 198.1)^{16.08}$ \\
\hline
\end{tabular}
\caption{The atmosphere model adopted from Ref.\ \cite{Mahdawi:2018euy}. Here 
$z$ is the altitude, 
$z_{\rm min}$ ($z_{\rm max}$) is the minimum (maximum) altitude of each layer,
and $\rho_{\rm atm}(z=0)=1.225\times 10^{-3}$ g/cm$^{3}$ 
is the atmosphere density at the sea level. 
The atmosphere density is taken to be zero
above 84.8 km.}
\label{tab:atmosphere}
\end{table}

For the SI case, we consider both N and O elements in the atmosphere that scatter with DM. The mass fractions of N (O) in the atmosphere is 78.08\% (20.94\%). 
For the SD case, we only consider N in the atmosphere that can scatter with DM through SD interaction. We neglect the contribution of O 
for the CSR upper boundary in the SD case, because the natural abundance of $^{17}$O (the O isotope that has spin) is 0.04\% \cite{lodders1998planetary}.

\section{Thermalization efficiency}
\label{sec:thermal}

The effects of the thermalization efficiency, 
the ratio between the measured energy in XQC 
and the DM nuclear recoil energy, 
have been recently investigated in 
Ref.\ \cite{Mahdawi:2018euy}, 
for the parameter space of $m_\chi\in\left[0.01,100\right]$ GeV 
and 
$\sigma_{\chi p}\in\left[10^{-31},10^{-21}\right]$ cm$^2$. 
In this section, we study the effects of the thermalization efficiency 
on the XQC contour in 
the entire parameter space 
spanned by the DM-proton 
cross-section and the DM mass.

Fig.\ (\ref{fig:XQC-thermal}) 
shows the 
XQC exclusion region 
in the SI case where $n=0$ and $f_\chi=1$, 
for three different thermalization efficiencies: 
$\epsilon=1$, 
$\epsilon=0.1$, 
and $\epsilon=0.02$. 
These three thermalization efficiencies 
have been recently suggested by Ref.\ \cite{Mahdawi:2018euy}, 
since the experimentally measured value 
is not available in the energy range 
of interest for the XQC detector. 
As shown in Fig.\ (\ref{fig:XQC-thermal}), as $\epsilon$ decreases, 
the left boundary of XQC exclusion reigon 
shifts to the right, 
and the right boundary shifts to the {left}. 
The effects of the thermalization efficiency 
are less significant on the upper and lower 
boundaries of the XQC exclusion contours. 
We note that the difference at the lower boundary 
in the mass range of $\sim(1-100)$ GeV
between our calculation and Ref.\ \cite{Mahdawi:2018euy} 
is due to the effects of DM penetrating the aluminum body of the rocket 
considered in Ref.\ \cite{Mahdawi:2018euy}.

\begin{figure}[tbp]
    \centering
    \includegraphics[width=0.45\textwidth]{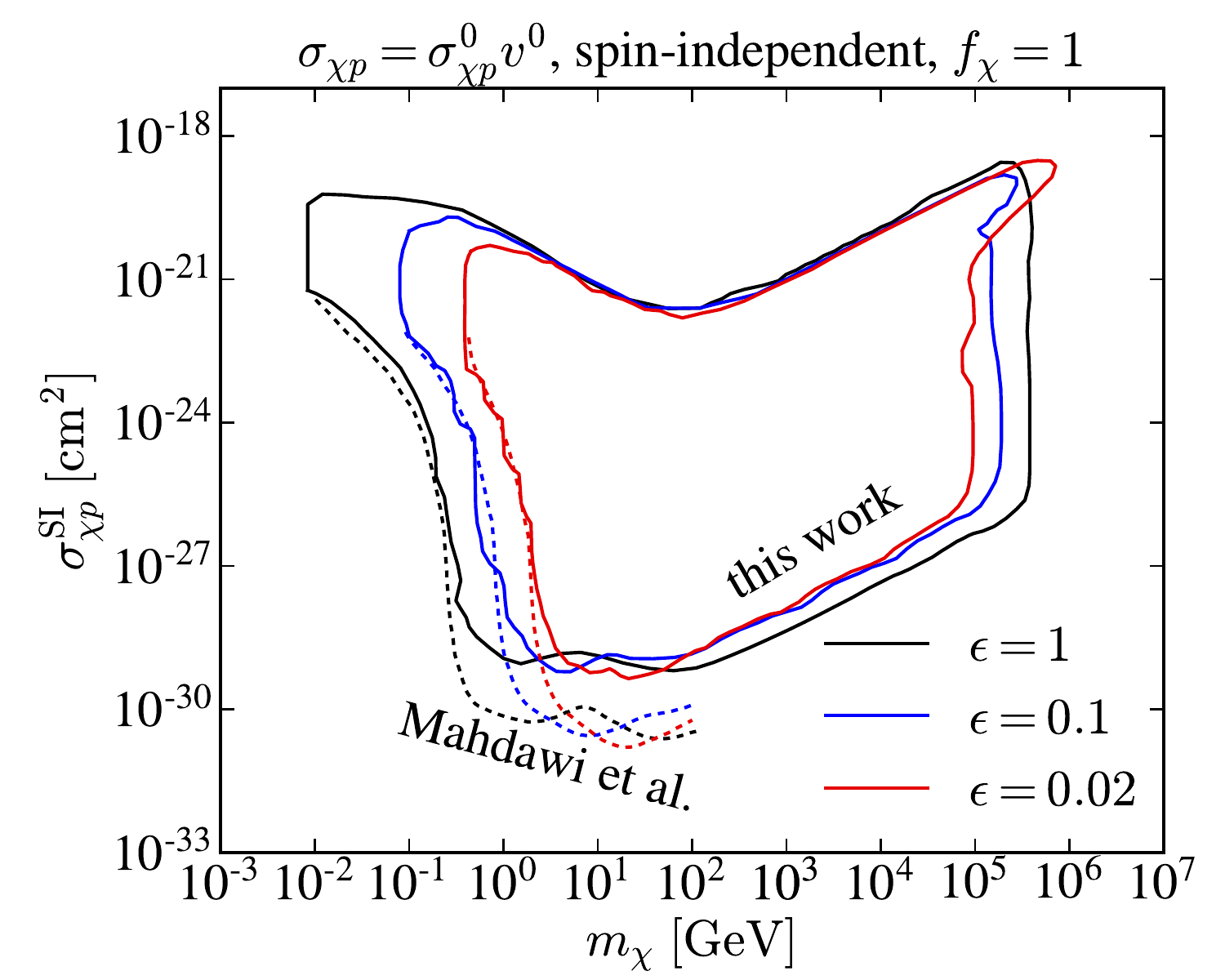}
    \caption{XQC exclusion region 
    in the SI case where $n=0$ and $f_\chi=1$, 
    for different thermalization efficiencies: 
    $\epsilon=1$ ({\it black-solid}), 
    $\epsilon=0.1$ ({\it blue-solid}), 
    and $\epsilon=0.02$ ({\it red-solid}). 
Results from Ref.\ \cite{Mahdawi:2018euy} 
(Mahdawi et al.)
    are shown in dashed lines: 
$\epsilon=1$ ({\it black-dashed}), 
$\epsilon=0.1$ ({\it blue-dashed}), 
and $\epsilon=0.02$ ({\it red-dashed}). } 
    \label{fig:XQC-thermal}
\end{figure}


\bibliography{ref.bib}{}
\bibliographystyle{JHEP}

\end{document}